\documentclass{article}
\usepackage{hyperref}
\usepackage{subfig}
\usepackage{natbib}
\usepackage{authblk,booktabs}
\usepackage{graphicx}

\usepackage{amsmath}
\usepackage{amssymb,amsfonts}
\usepackage[left=1.2in,top=1.1in,right=1.2in,bottom=1.1in]{geometry}
\usepackage[noabbrev]{cleveref}

\DeclareMathOperator{\ddt}{\dfrac{d}{d t}}

\newcommand{\spardif}[2]{\ensuremath{\frac{\partial #1}{\partial\,#2}}}

\begin{document}

 \title{Dynamical Analysis of Blocking Events: Spatial and Temporal Fluctuations of Covariant Lyapunov Vectors}
 \author[1,2]{Sebastian Schubert}
 \author[2,3,4]{Valerio Lucarini}
 \affil[1]{IMPRS - ESM, MPI f\"ur Meteorologie, University of Hamburg, Hamburg, Germany, Email: sebastian.schubert@mpimet.mpg.de}
 \affil[2]{Meteorological Institute, CEN, University Of Hamburg, Hamburg, Germany}
 \affil[3]{Department of Mathematics and Statistics, University of Reading, Reading, United Kingdom}
 \affil[4]{Walker Institute for Climate System Research, University of Reading, Reading, United Kingdom}
 \date{First Draft: November 2015\\
This Draft: January 2016}
\maketitle
\begin{abstract}
One of the most relevant weather regimes in the mid-latitudes atmosphere is the persistent deviation from the approximately zonally symmetric jet stream to the emergence of so-called blocking patterns. Such configurations are usually connected to exceptional local stability properties of the flow which come along with an improved local forecast skills during the phenomenon. It is instead extremely hard to predict onset and decay of blockings.
Covariant Lyapunov Vectors (CLVs) offer a suitable characterization of the linear stability of a chaotic flow, since they represent the full tangent linear dynamics by a covariant basis which explores linear perturbations at all time scales. Therefore, we  assess whether CLVs feature a signature of the blockings.
As a first step, we examine the CLVs for a quasi-geostrophic beta-plane two-layer model in a periodic channel baroclinically driven by a meridional temperature gradient $\Delta T$. An orographic forcing enhances the emergence of localized blocked regimes. We detect the blocking events of the channel flow with a Tibaldi-Molteni scheme adapted to the periodic channel.
When blocking occurs,  the global growth rates of the fastest growing CLVs are significantly higher. Hence, against intuition, the circulation is globally more unstable in blocked phases.
Such an increase in the finite time Lyapunov exponents with respect to the long term average is attributed to stronger barotropic and baroclinic conversion in the case of high temperature gradients, while for low values of $\Delta T$, the effect is only due to stronger barotropic instability. In order to determine the localization of the CLVs we compare the meridionally averaged variance of the CLVs during blocked and unblocked phases. We find that on average the variance of the CLVs is clustered around the center of blocking. These results show that the blocked flow affects all time scales and processes described by the CLVs.   
\end{abstract}

\section{Introduction}
The study of weather regimes in the atmosphere is a key topic in meteorology and geosciences. In particular, blocking highs have been early on identified as persistent, large scale deviations from the zonally symmetric general circulation \citep{Rex1950,Baur1947}. Traditionally, the detection and description of these events employs objective indicators based on pressure anomalies in the  atmosphere obtained from observational data or output of general circulation models \citep{Lejenaes1983,TIBALDI1990,Schalge2011}. Such blocking events and related large scale weather regimes provide an important contribution to the low frequency variability of the atmosphere. In particular, one can interpret the mid-latitude atmosphere as jumping between a zonal regime and a blocked regime, or, more in general, a regime where long waves are strongly enhanced  \citep{Benzi1986,Sutera1986,Molteni1988,Ruti2006}. One needs to remark that the so-called bimodality theory and the analyses which have confirmed - at least partially - its validity have been criticized in the literature, see e.g. \citet{Nitsche1994} and \citet{Ambaum2008}. In \citet{Charney1979,Charney1980}, it was speculated that the existence of multiple stationary equilibria in simple models of the atmospheric circulation is the root cause for weather regimes. In their investigation of a highly truncated quasi-geostrophic (QG) models, several stationary states exist due to an orographic forcing. Different weather regimes are then associated with the neighborhood of the various stationary states. Contrary to this theory of multiple equilibria, it was found that in less severely truncated models, which  adopted realistic forcings, stationary states are far away from the attractor and/or only one stationary state exists \citep{Reinhold1982a,Tung1985a,Speranza1988}. In a recent contribution by \citet{Faranda2015}, a different paradigm is instead proposed: blocking events are seen as close returns to an unstable fixed point in a suitably defined reduced space describing the large scale dynamics of the atmosphere.

When considering high-dimensional chaotic dynamics, we have to look at the problem of the possible existence of weather regimes by looking at the properties of the invariant measure supported on the attractor of the system. A possible way to revisit the idea of transitions between atmospheric regimes is based on looking at the switching between  the neighborhood of unstable periodic orbits \citep{Gritsun2013}. We remind that unstable periodic orbits provide an alternative way to reconstruct the properties of the attractor of a chaotic dynamical systems (Cvitanovic and Eckhard 1991). Also, heteroclinic connections between unstable stationary states were found in a highly truncated barotropic model \citep{Crommelin2003}. In models with higher complexity \textit{leftovers} of these structures are found and correlate with transitions between different weather regimes \citep{Kondrashov2004,Sempf2007}. In a reduced model phase space, this allows for identifying different dynamically stable weather regimes and less stable transitions paths between them \citep{Tantet2015}.

In this paper, we take inspiration from the classical point of view on the dynamics of blocking, which focuses on the analysis of the linear instabilities of low-order models, but here we we consider more Earth-like - at least, qualitatively - background turbulent atmospheric conditions. While the attractors we consider are strange geometrical objects, we follow a mathematical approach such that we are able to stick to the investigation of linear stability properties, which allows for a relatively easy interpretation of the underlying physical mechanisms. Ever since \citet{Lorenz1963}, it is clear that linear stability is a measure of predictability of the atmosphere. Therefore, the difficulty of predicting - in time - the onset and decay of weather regimes and their persistence should be reflected in local stability properties.  The analysis of optimal linear perturbations indicated that the leading optimal perturbation localizes where blocking occurs \citep{Buizza1996}. In a study by \citet{Frederiksen1997} normal modes for a time varying basic state were investigated. \citet{Naoe2002} found that - in contrast to the baroclinic instability - the emergence of blocking  events can not be explained by linear perturbations of fixed states of the atmosphere, instead non-linear processes have to be included.

Our approach to this problem will be based on investigating blocking events using Covariant Lyapunov Vectors (CLVs). These vectors form a norm-independent and covariant basis of the tangent linear space \citep{Ruelle1979,Eckmann1985,Trevisan1998,Ginelli2007}. The long-time average of the growth rates of the CLVs  give the Lyapunov exponents (LEs), see discussion in \citet{Froyland2013} and \citet{Vannitsem2015}. Note that by spanning the tangent space of the attractor, CLVs allow in principle a precise calculation of the response operator to an arbitrary perturbation of a dynamical system \citep{Lucarini2014,Lucarini2011,Ruelle2009}. CLVs provide a powerful method for characterizing the properties of weather regimes. First, they are a first order representation of the dynamics around a fully non-linearly evolving background state, so that no simplifying hypotheses are made on the dynamics. Second, they are a generalization of the normal mode instabilities of basic states of the atmosphere, so that it is still possible to use all the machinery of linear ordinary differential equations. Taking these points into consideration, it is suggestive to consider CLVs as a superior choice over other orthogonal, hence norm-dependent Lyapunov vectors \citep{Legras1996}. 

Previously, we have investigated CLVs in a quasi geostrophic two layer model in a periodic channel \citep{Schubert2015}. In that work, we addressed how the average energy and momentum transports of the CLVs are related to their growth and decay in respect to the background state and how they explain the variance of the background state. Moreover, we provided a bridge between the growth rate of the CLVs and the physical mechanisms responsible for the variability of the quasi-geostrophic flow, namely the barotropic and baroclinic conversion, by a detailed analysis of the Lorenz Energy cycle of each CLV. We note that our focus was exclusively on the long-term properties of the flow, of its CLVs, and of the corresponding LEs. 

In this paper, we are concerned with weather regimes in the background state, hence we will study the fluctuations of the CLVs and of the finite-time LEs. The rationale of our study is then the following. Using the classical Tibaldi-Molteni scheme blocking detection, we will determine when the flow is unblocked and when/where the flow switches to a blocked state \citep{TIBALDI1990}.
We will then address two questions.
\begin{enumerate}
\item Is there a systematic signature of blocked phases in the growth rates of the linear perturbations? Is there a systematic change in the energetics of the flow? 
\item Is the occurrence of blocked phases linked to the presence of specific patterns for the CLVs and to their localization in the physical space?
\end{enumerate}
Note that the second question is different from the average localization of the CLVs investigated in \citet{Szendro2008}.
In order to address these questions, we have extended the model of our previous study with an orographic forcing, following \citep{Charney1980}. As in our previous study, the model is baroclinically driven by introducing a relaxation meridional temperatrue gradient $\Delta T$ and dissipates energy via Ekman pumping, which parameterizes the effect of the planetary boundary layer. The orography in our investigation is a Gaussian bump in the middle of the domain with horizonal scale of $O(1000)$ $km$. 
We explore the sensitivity of the problem by considering multiple setups featuring different heights of the gaussian bump and different values of $\Delta T$. The various setups all exhibit chaotic conditions with many positive LEs. 

We find that blocking increases with a higher meridional temperature gradient $\Delta T$ and is additionally enhanced by orography, which, by breaking the zonal symmetry, contributes as a catalyst to the process of a phase lock mechanism that allows standing perturbations to grow, as envisioned in \citet{Benzi1986}.
The spatial variance of the CLVs is dominantly located around the region where blocking occurs when compared to the average variance during unblocked phases. Furthermore, the growth rates of the fastest growing CLVs are higher during blocked phases, pointing at the fact that the system has globally a lower predictability during blocked phases, possibly as a result of the difficulty of predicting when the onset and decay of the blocking events. The observed increased instability suggests also that the local dimension of the attractor is higher during blocking. We explain the changed growth behavior by using a generalization of the Lorenz energy cycle between the CLVs and the background state introduced in \citet{Schubert2015}. We find that for high values of $\Delta T$ the increased instability is dominantly caused by an increased input of energy to the CLVs by baroclinic and barotropic conversions, while for weakly baroclinic flows the intensification of barotropic instability is the only active mechanism.
We speculate that these results hint at a possible definition of blocking by taking into account the properties of all CLVs at a particular time.

The structure of the paper  can be summarized as follows. In \cref{sec:experimentalsetup}, we describe our experimental set-up, by sketching the formulation of the QG model, the basic properties of CLVs, and the blocking detection algorithm.
In \cref{sec:III}, we present our results on the properties of blocked vs unblocked phases, discussing the properties of the fluctuations of LEs and CLVs and investigating the sensitivity of our results to changes in the forcing and in the orography. Finally, in \cref{sec:conclusion}{}, we give an outlook and summary and point the reader towards future work on these topics.

        \section{Experimental Setup}\label{sec:experimentalsetup}
\subsection{The Model}\label{sec:model}
\begin{figure*}[ht!]
     \centering
     \subfloat[][Vertical Structure]{\includegraphics[width=0.5\textwidth]{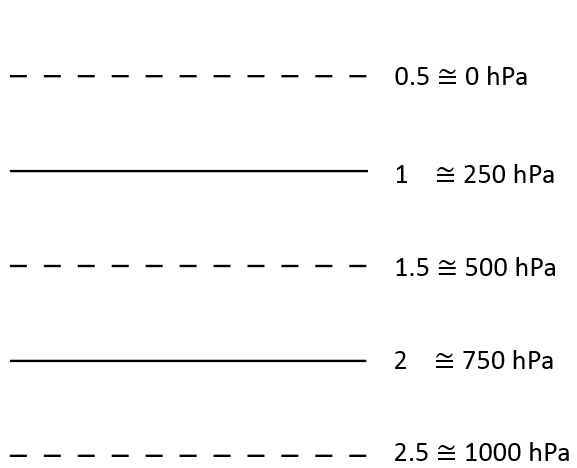}\label{fig:layers}}
     \subfloat[][Horizontal Domain]{\includegraphics[width=0.5\textwidth]{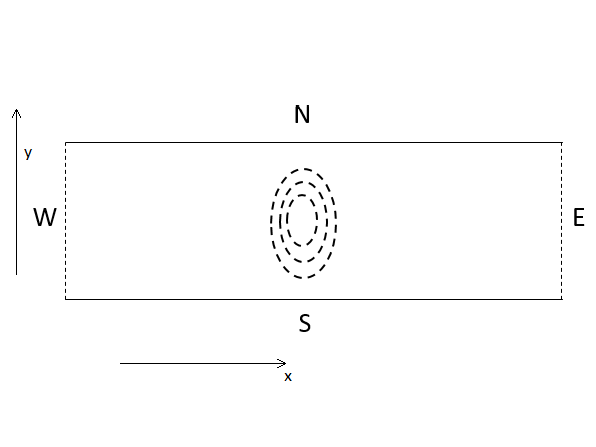}\label{fig:horizontaldomain}}
     \caption{The domain of the QG model. The right panel shows the geometry of the horizontal domain and in dashed lines the position of the orography implemented as a forcing on $\omega_{2.5}$\label{fig:domain}}
\end{figure*}

We conduct our investigation with a two layer model of the mid-latitudes featuring the basic large scale baroclinic and barotropic processes of the atmosphere.  Previously, we have used the same model without orography to obtain the CLVs in \citet{Schubert2015}. The model is a spectral version of the classical model introduced by \citet{Phillips1956} extended by an orographic forcing. This type of model was earlier used to investigate blocking and stationary states by \citet{Charney1980}. The horizontal domain is rectangular $\left(x,y \right)\in [0,L_x]\times [0,L_y].$  It is periodic in the x-direction and a no-flux condition is imposed at $y=0,\pi$ (see \cref{fig:horizontaldomain}). In the vertical, two layers are resolved (see \cref{fig:domain}). We solve the quasi geostrophic vorticity equation at the pressure levels $p_2=750 hPa$ and $p_1=250 hPa$ level and the thermodynamic equation at the pressure level $p_{1.5}=500 hPa$ (see \cref{fig:layers}).
The model is forced by a newtonian cooling towards a zonally symmetric temperature profile $${T_{e}}=\frac{\Delta T}{2}\cos\left(\frac{\pi y}{L_y}\right).$$ Small scale interactions are parameterized through eddy diffusion $k_h\Delta^2\psi$. The interaction with the boundary layer due to Ekman pumping and orography $$h_{GB}=h_0 e^{-\left(\frac{x-L_x}{\sigma_x}\right)^2-\left(\frac{y-L_y}{\sigma_y}^2\right)}$$ is parameterized by imposing in the lower layer vertical p-velocity $\omega_{2.5}=\frac{\Delta p}{f_0}\ 2r \Delta\psi_2 - \frac{\Delta p}{H} J(\psi_2,h_{GB})$, see, e.g. \citet, where $H$ is the height of the atmosphere ($7.3 km$). Note that for this implementation of orography we have to ensure that $h/H$ at least smaller than 1. At the top of the atmosphere  $\omega_{0.5}$ is zero. The system is solved in terms of the geostrophic streamfunction $\psi $ and the temperature $T$, see \citet{Holton2004}. 
\begin{subequations}
\begin{align}
\begin{split}
\ddt \Delta\psi_{1} =&-J\left(\psi_1,\Delta\psi_1 + f_0 + \beta y\right) + f_0\frac{\omega_ {{1.5}}-\omega_ {{0.5}} }{\Delta p}\\&+k_h\Delta^2\psi_1 \label{eq:lay1}
\end{split}\\
\begin{split}
\ddt \Delta\psi_{2} =&-J\left(\psi_2,\Delta\psi_2 + f_0 + \beta y\right) + f_0\frac{\omega_ {{2.5}}-\omega_ {{1.5}} }{\Delta p}\\&+k_h\Delta^2\psi_2\label{eq:lay2}
\end{split}\\
\ddt T =&-J\left(\psi_M,T\right)+ S_p \omega_{1.5} +r_R \left(T_{e}-T\right)+\kappa\Delta T \label{eq:thermo}
\end{align}
\end{subequations}
Where \cref{eq:lay1,eq:lay2} are the QG vorticity equation and \cref{eq:thermo} is the QG thermodynamic energy equation. For more details on the model itself, we refer to the model description in \citet{Schubert2015}. The adimensionalization is performed according to \cref{tab:parameters}. We introduce also a modified stability parameter $S=S_p\frac{R \Delta p}{2f_0^2}$. The adimensional equations are then the following. We define a barotropic $\psi_M=\frac{\psi_1+\psi_2}{2}$ and baroclinic streamfunction $\psi_T=\frac{\psi_1-\psi_2}{2}$,
the latter being proportional to the temperature $T=\frac{2f_0}{R}\psi_T$.
We then obtain:
\begin{align}\label{eq:eom}
\begin{split}
\ddt\Delta\psi_M =&-J(\psi_M,\Delta \psi_M + \beta y) - J(\psi_T,\Delta\psi_T)\\&-r\Delta(\psi_M-\psi_T)+k_h \Delta^2\psi_M \\
&+\frac{1}{2}J(\psi_M-\psi_T,h_{GB})\\
\ddt\,\,\Delta\psi_T =&-J(\psi_T,\Delta \psi_M + \beta y) - J(\psi_M,\Delta\psi_T)\\
&+ r\Delta(\psi_M-\psi_T)+k_h \Delta^2\psi_T \\
&- \frac{1}{2}J(\psi_M-\psi_T,h_{GB})+\omega
\\\ddt\,\, \psi_T\,\, \,\,=&-J(\psi_M,\psi_T) + S \omega + r_R \left(\psi_{Te}-\psi_T\right) \\
&+ \kappa \Delta \psi_T 
\end{split}
\end{align}

\begin{table}
\centering
\begin{tabular}{llllll}
\toprule
Variables, Operators & Symbol &Unit& Scaling & Value of& \\
\& Constants & & &Factor  & Scaling Factor&\\
\midrule
Streamfunction & $\psi$  & $m^2 /s$   & $L^2f_0$&$10^{10}/\pi^2$  &\\
Temperature & $T$           &$K$   & $2f_0^2L^2/R$        &$705.97$  &\\
Velocity        & $\textbf{v}$&$m/s$  & $Lf_0$   &$10^3/\pi$  &\\
Laplace Operator & $\Delta$&$1/m^2$  & $1/L^2$   &$\pi^2/10^{14}$  &\\
Vertical p-Velocity & $\omega$&$Pa/s$  & $\Delta p f_0$   &$0.01$  &\\
Jacobian & $J(\cdot,\cdot)$&$1/m^2$  & $1/L^2$   &$\pi^2/10^{14}$  &\\
\midrule
\midrule
Parameters & Symbol & Dimensional & Unit & Scaling&Non-dimensonal\\
&&Value&&Factor&Value\\
\midrule
Forced Meridional & $\Delta T$  & $40 - 76$ & $K$  & $2f_0^2L^2/R$ &$0.0567 - 0.1076$\\
Temperature Gradient &&&&\\
Height Of Mountain & $h_0$  & $1.48$, $2.96$, $4.44$ & $km$ & $H$ & $0.2,\,0.4,\,0.6$\\
Width of Mountain & $\sigma_{x}, \sigma_{y}$  & $1000,\, 2000 $ & $km$ & $L$ & $0.1\pi,\,0.2\pi$\\
Eddy-Heat Diffusivity & $\kappa$ & $10^5$ &$m^2/s$& $L^2f_0$&$9.8696 \cdot 10^{-5}$\\
Eddy-Momentum Diffusivity & $k_h$ & $10^5$ &$m^2/s$&$L^2f_0$&$ 9.8696 \cdot 10^{-5}$\\
Thermal Damping & $r_R$ & $1.157\cdot 10^{-6}$ &$1/s$&$f_0$& 0.011\\
Ekman Friction & $r$ & $2.2016 \cdot 10^{-6}$ &$1/s$& $f_0$&0.022\\
Stability Parameter & $S$& $3.33 \cdot 10^{11}$ & $m^2$&$L^2$&0.0329\\
Coriolis Parameter & $f_0$ & $10^{-4}$& $1/s$&$f_0$&1\\
Beta & $\beta$ & $1.599 \cdot 10^{-11}$ & $1/(ms)$ &$f_0/L$&0.509\\
Aspect Ratio &$ a $ & $0.6896 $ & 1 & - & 0.6896 \\
Zonal Length & $L_x$ & $2.9\cdot 10^7$ & $m$&L& $\frac{2\pi}{a}$ \\
Meridional Length & $L_y$ & $10^{7}$ & $m$&L&$\pi$\\
Height Of Atmosphere & $H$ & $ 7.4$ & $km$\\
Specific Gas Constant &$R$ &$287.06$&$J/(kg K)$&$R/2$&$2$\\
Pressure &$\Delta p$ &$500 hPa$&$N/m^2$&$\Delta p$ & $1$\\
\bottomrule
\end{tabular}
\caption{Parameters and Variables used in this model and the respective adimensionalization scheme. Note that the scales for time and length are $t=10^4 s = 1/f_0$ and $L=\frac{10^7}{\pi} m$\label{tab:parameters}}
\end{table}
The orography $h_{GB}$\ is an idealized Gaussian bump designed to resemble loosely the scales of the Rocky Mountains placed in the middle of the horizontal domain. Hence, $\sigma_x=1000\, km$ and $\sigma_y=2000\, km$.
 We integrate the equations in spectral space using the following decomposition.
\begin{align}
\begin{split}
 &\displaystyle\psi(x,y,t)= \sum_{k,l=1}^{N_x,\, N_y} \left(\psi^r(k,l,t)\cos\left(a k x\right) \right.\\
 &\left.+\psi^i(k,l,t)\sin\left(a k x\right) \right)\sin\left(l y\right) + \sum_{l=1}^{N_y} \psi^r(0,l,t) \cos\left(l y\right) \\
\end{split}
\end{align}
The spectral cutoff is in the zonal direction at $N_x=10$ and in the meridional direction at $N_y=12$.
The total dimension of the model phase space is $2N_y(2N_x+1)=504$. This resolution is rather coarse but nevertheless still sufficient for capturing the large scale structure that we are interested in, see \citet{Schubert2015}. We perform a spin up run of 30 years. All results will be based on a time series of 31 years. 
We investigate three different mountain heights $h_0$ (1.48 km, 2.96 km and 4.44 km) and four different meridional temperature gradients $\Delta T$ (40 K, 50 K, 66 K and 76 K). This ensures the investigation of different states of large scale turbulence and the assessment of the impact of orography. In control runs, the experiments are done without orography.  The implemented 4\(^{th}\) order Runge-Kutta-Scheme uses a fixed time step of $2.77$ hours (1 adimensional time unit) except for the highest $\Delta T=76 K$, where, instead, we choose a time step of  $1.385$ hours (0.5 adimensional time units). The analysis of the data is sampled every $2.77$ hours. For our analysis, we investigate a total time series of 31 years. For this we consider additionally 15 years as spin up before and after the 31 years in order to compute the CLVs. 
\subsection{Covariant Lyapunov Vectors}\label{sec:CLV}

CLVs are a powerful tool for investigating the tangent linear model of a dynamical system. We have previously summarized theory of CLVs in \citet{Schubert2015} and how they can be obtained via the algorithm of \citet{Ginelli2007}.  

Let us briefly report on the main properties of the CLVs and their importance for characterizing the dynamics of small perturbations. Summarizing, they represent the covariant directions of expansion and decay in the linear tangent space which grow on average with the LEs, see \citet{Legras1996}. If the background state is a simple fixed point, they reduce to the classical normal modes of stationary solutions, see \citet{Wolfe2007}. For periodic background, they coincide with the Floquet vectors \citep{Floquet1883,Samelson2001a,Wolfe2006,Wolfe2008}. These properties make them an ideal choice to describe the growth (in linear approximation) of actual physical disturbances, since the classical Lyapunov vectors (Gram-Schmidt vectors) are orthogonal and can therefore not describe the covariant evolution of nearby trajectories of the background flow, see \citet{Pazo2010,Kuptsov2012}. Following \citet{Szendro2008}, we also expect that in the specific case the system is prepared in such a way that a localized perturbation breaks otherwise symmetric boundary conditions, one can observe localization phenomena (not exclusively in the vicinity of the perturbation) for the CLVs of the system. This last property is extremely attractive for the problem studied in this paper.
 
The evolution of CLVs describes how infinitesimal perturbations added to the background flow $\textbf{x}_B$ change in time. From a dynamical system point of view \cref{eq:eom} can be written in the following form.
\begin{equation} \label{eq:dyn}
 \ddt \textbf{x} = f(\textbf{x})
\end{equation}
An infinitesimal perturbation $\textbf{v}$ added to $\textbf{x}_B$ solves the following equation which is obtained by conducting a first order expansion of \cref{eq:dyn}. 
\begin{equation}\label{eq:tanglinear}
  \ddt \textbf{v}_j(t) = \sum_i\spardif{f_j}{x_i}\left(\textbf{x}_B(t)\right)\ \textbf{v}_i(t) =: \sum_i\mathcal{J}_{ji}\left(\textbf{x}_B(t)\right)\ \textbf{v}_i(t).
\end{equation}
A normalized, covariant basis for the solutions of this equation is the CLVs $\left\{\textbf{c}(x_B(t))_j\right\}_{j=1..n}$ (n is here 504). Let us explain what this means more precisely. For this, we need besides the normalized vectors $\textbf{c}_j\left(t\right)$ also the corresponding time series of growth rates $\lambda_i(t).$ Their average is equal to the jth LE $\lambda_j=\lim_{T\rightarrow \infty}\frac{1}{T}\int_0^T dt \lambda_j(t)$. If we pick a solution $\textbf{v}$ of equation \ref{eq:tanglinear} at time $t$ with the initial condition $\textbf{v}(t_0)=\textbf{c}_j(t_0)$ then $\textbf{v}(t)$ has the following form.
\begin{equation}\label{eq:clv_ts}
\textbf{v}(t)=e^{\int_{t_0}^t\, dt' \lambda_j(t')}\, \textbf{c}_j(t)
\end{equation}
Note that the normalized CLVs $\textbf{c}_j(t)$ solve the following slightly altered equation.
\begin{equation}\label{eq:norm_clv}
\dot{\textbf{c}}_j(t)=\mathcal{J}(\textbf{x}_B(t))\textbf{c}_j(t)-\lambda_j(t)\textbf{c}_j(t)
\end{equation}
Imagine now, we chose an arbitrary initial condition $x_0$ at time $t_0$, hence a superposition of possibly all CLVs $$\textbf{x}_0=\sum_j \textbf{c}_j(t_0)A_j.$$ Consequently, the solution $\textbf{x}$ for \cref{eq:tanglinear} with $\textbf{x}(0)=\textbf{x}_0$ has the following form in the basis of the CLVs. $$\displaystyle\textbf{x}(t)=\sum_j e^{\int_{t_0}^t\, dt' \lambda_j(t')}\, \textbf{c}_j(t )A_j $$
This means the expansion into the CLVs allows it to investigate also very slow growing linear perturbations without interference of the fast growing directions. Note that this is particularly interesting for applications in data assimilation, where models often feature multi scale interactions \citep{Pazo2010}.
These properties are unique to CLVs. Other Lyapunov vectors, e.g. Gram-Schmidt vectors, see \citet{Kuptsov2012},
are not solutions of  \cref{eq:tanglinear}.
Hence, CLVs describe to first order solutions of \cref{eq:dyn} that are nearby to $\textbf{x}_B$. Consequently, they have a straightforward physical interpretation and we can obtain meaningful transports and feedbacks of the CLVs connected to the background $\textbf{x}_B$ \citep{Schubert2015}.

For further details on the algorithm to obtain\ CLVs we refer to \citet{Ginelli2007,Kuptsov2012} and our previously mentioned previous description in \citet{Schubert2015}.

\subsection{Blocking Detection}\label{sec:blockingdetection}
We describe briefly the adapted Tibaldi-Molteni scheme \citep{TIBALDI1990} for detecting blocking highs in our model.
Since the model is spectral, we are using a Fast-Fourier-Transform-algorithm to transform the spectral fields to a $\left[64 \times 32\right]$ grid (64 grid points in the x direction, 32 grid points in the y direction).
We will consider only blocking in the barotropic streamfunction $\psi_P$, since it is the best representation of the 500 hPa layer in our model discretization. In order to detect blocking high anomalies, we study the occurrence at some longitude of reversals in the direction of the zonal wind with respect to normal conditions. We construct the average zonal wind in the northern and southern sector by constructing the quantities $u_N(x,\Delta)=-\delta \psi_N(x,\Delta) / (y_N-y_0)$ and $u_S(x,\Delta)=-\delta \psi_S(x,\Delta)/ (y_0-y_S)$. where $\delta\psi_N(x,\Delta)=\psi_P(x,y_N+\Delta)-\psi_P(x,y_0+\Delta)$ and $\delta\psi_S(x,\Delta)=\psi_P(x,y_0+\Delta)-\psi_P(x,y_S+\Delta)$.

A blocking event occurs, if at a particular coordinate x $u_S$ is negative and $u_N$ is sufficiently positive and lasts at least two days. We also allow for a deviation $\Delta$ from the chosen y-coordinates.
Summarizing this we have the following criteria.\begin{align}\label{eq:TibaldiMolteni}
\begin{split}
&u_N(x,y_0+\Delta)>9\, m/s\\
&u_S(x,y_0+\Delta)<0\, m/s\\
&y_N=8437\, km; \,\,y_0=6250\, km;\,\, y_S=4375\, km\\
&\Delta=(-940\, km\,\, 0\, km\,\, 940\, km)
\end{split}
\end{align}
A word of caution is needed at this point, since the Tibaldi-Molteni index was originally developed for a spherical geometry and considered either observational data or more realistic data taken from GCMs. We still think the use of this index is meaningful, because of its straightforward interpretation and because it describes the presence of an non-zonal deviation from the usually fluctuating, but zonally symmetric jet stream. We will show that the detected blocking events are indeed meaningful, hence a blocked and an unblocked weather regime can be determined (see the following discussion in \cref{sec:blockingpatterns}).
\section{Results}\label{sec:III}
\subsection{Blocking Events}\label{sec:blockingpatterns}
\begin{figure}[h!]
\centering
 \includegraphics[width=\textwidth]{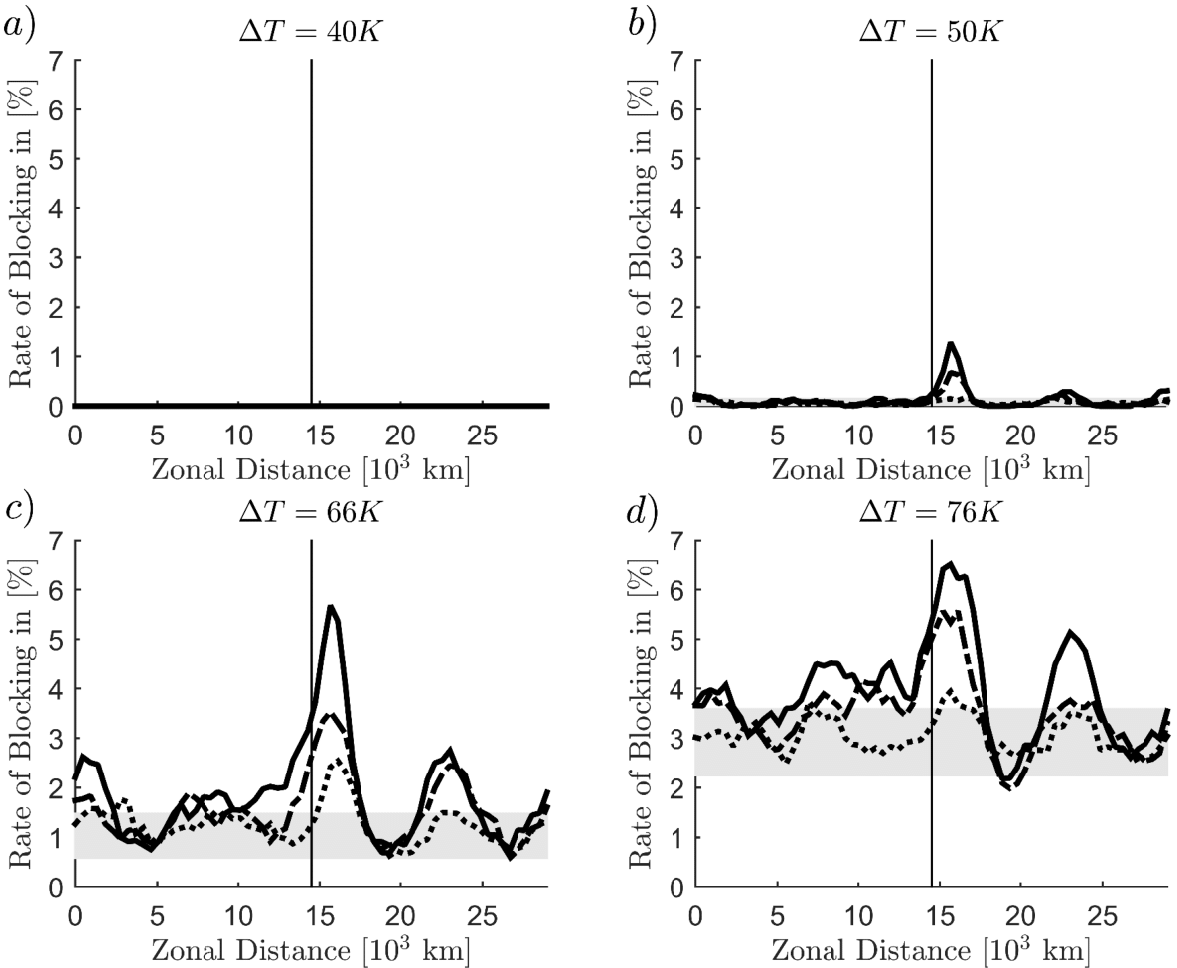}
 \caption{The number of blocked days is the highest behind the peak of the orography. The vertical black line indicates the peak of the orography. The panels show the different values of $\Delta T$, the different heights $h_0$ are indicated by the dotted line (1.48 km), the dash-dotted line (2.96 km) and the solid line (4.44 km). Downstream two secondary maxima can be identified. The x axis indicates the x coordinate where we detect blocking. The y axis shows in percent the frequency of blocking. The grey shaded area shows the range of the blocking rate along the x direction without orography. \label{fig:blockingrate}}
\end{figure}


\begin{figure}[ht!]
\centering
 \includegraphics[width=\textwidth]{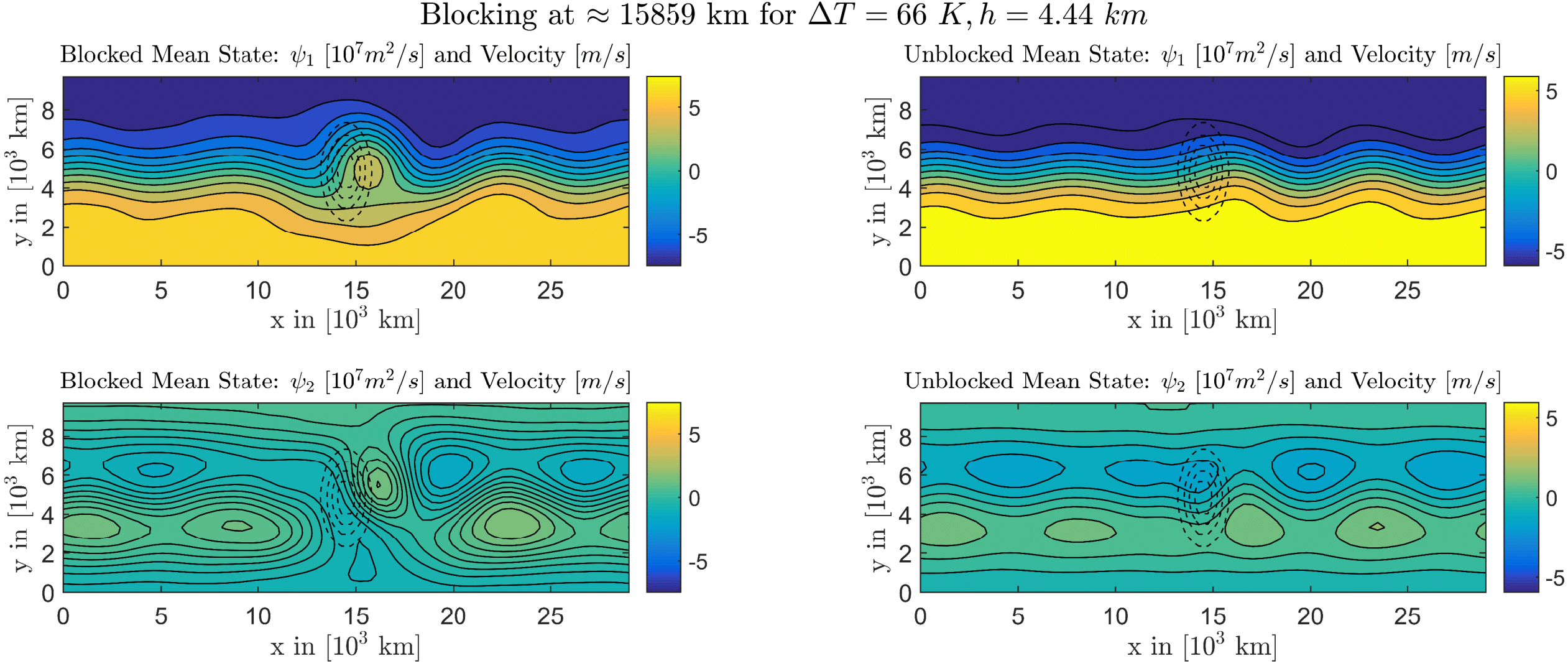}
 \caption{As an example of the observed blocking events, the mean blocked state at $x=15859\, km$ and the unblocked state for $\Delta T =66\, K$ and $h=4.44\, km$ is shown. The left panels show the averaged streamfunction during blocking. Note that, the streamfunction in our QG model should be compared to the  stratified streamfunction in models without the hydrostatic balance. This means the upper layer streamfunction $\psi_1$ and the lower streamfunction $\psi_2$ determine the geostrophic velocities $\textbf{v}_{1/2}=\left(u_{1/2},v_{1/2}\right)=\left(-\partial_y\psi_{1/2},\partial_x \psi_{1/2}\right)$. The ageostrophic velocity can then be obtained via the QG momentum equations. The right panels show the average streamfunction for unblocked periods. The upper panels show the upper layer, the lower panels show the lower layer. The dashed lines show the position of orography. The blocking is affecting the flow only locally, since away from the blocking the flow is  the mean unblocked flow. We get similar results for blockings at different x coordinates and different values of $\Delta T$ and $h_0$.}\label{fig:block35}
\end{figure}

Let us first look at the blocking rate. Comparing the different setups with the control runs (without orography), we can also assess the impact of the orography on the blocking. For reasons of symmetry, in absence of orography, the statistics of blocking does not depend on x.

The blocking rate (see \cref{fig:blockingrate}) kicks off when $\Delta T$ is larger than 50 K (even without orography). The orographic forcing creates two to three local maxima in the blocking rate downstream of the peak of orography. 
These maxima intensify for higher $h_0$ and for higher $\Delta T$, but  within the range of values considered here the impact of $h_0$ reduces for higher $\Delta T$. 

 As mentioned before, two main configurations of the flow are identified. This can be further  substantiated by considering the mean states of the unblocked and blocked flow (e.g. $\Delta T=66 K$ and $h_0=4.44 km$ in \cref{fig:block35}). In this way, we treat the unblocked and blocked phases as separate weather regimes and determine the "climate" of the two, respectively. The blocked regime can be further divided into "sub regimes" by considering the mean state over the blocked phase and filtering for those parts of the blocked phase where only a chosen x coordinate is blocked. Since the days where blocking is present are relatively rare, the mean state  is computed over the complete time series is more or less identical to the mean state computed, taken over the days where no blocking is observed. Close to the blocked area a clear deviation from the zonal symmetric jet of the mean flow can be seen (the blocking high). We also see that the observed blocking is local and the regions far away from the blocked coordinate do not show large variations with respect to the average unblocked conditions. A first look at the mean unblocked flow shows that it is more zonally symmetric than the mean blocked flow. Nevertheless,  there is a non zonal disturbance with wave number four. This is caused by the presence of topographic Rossby waves induced by the orography \citep{Holton2004}. Note that the breaking of Rossby waves is intimately connected to the emergence of blocking events and the meandering of the jet fits roughly to the maxima of the blocking rate \citep{Berrisford2007}. 
The results shown in \cref{fig:block35} do not change significantly for the other setups and other locations, besides the shift of the blocking high to the corresponding x coordinate.

\begin{figure}[ht!]
\centering
 \includegraphics[width=\textwidth]{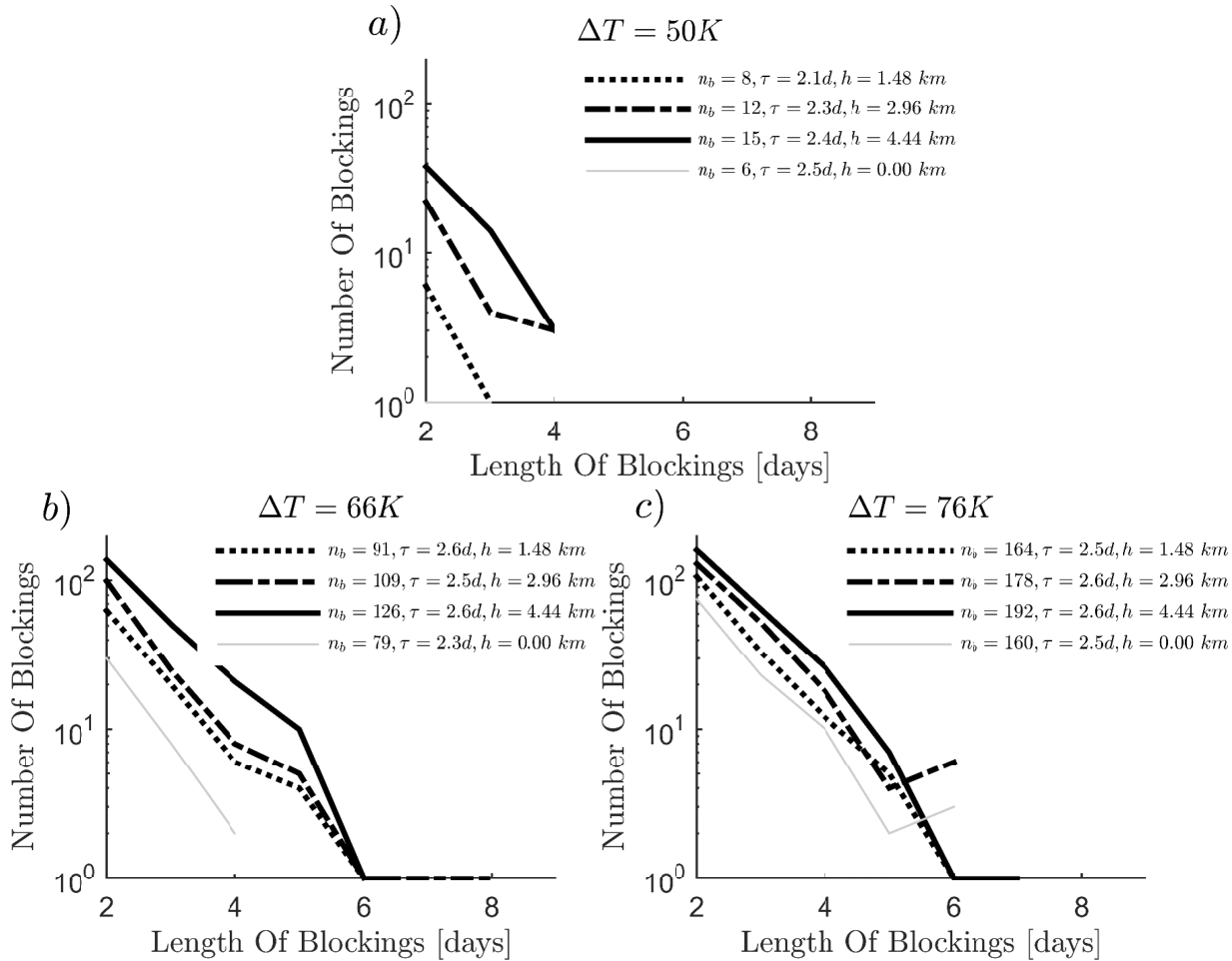}
 \caption{The distribution of blocking lengths at $x=15859\, km$ with orography in comparison to the control run without orography (light grey). The legends show also the lifetime $\tau$ of the blocking events and the number of blocked days per year $n_b$. The y-axis has a log scaling. The total length of the time series is 115705 days (31.7 years). \label{fig:blockinglength}}
\end{figure}

Let us turn our attention towards the number of blockings and their duration. We show the results for the position of the maximum of the blocking rate (see \cref{fig:blockinglength}), but the findings are similar at other x coordinates. 
The average blocking length (lifetime) is only marginally changed, whereas the total number of blockings is increased significantly by orography. 

For blocking rate and length (see \cref{fig:blockingrate,fig:blockinglength}), it appears that as discussed above, adding orography  creates preferential geographical locations for the occurrence of blocking. Looking at the global statistics, we have that, for a given value of $\Delta T$, the number of blocking events and the number of blocked days increase with $h_0$, even if the catalyzing effect of orographic disturbances is relatively weaker when $\Delta T$ is large enough. 
The effect of $\Delta T$ on the blocking rate and the number of blockings is also found in observations since the meridional temperature gradient is higher in winter and is associated with a higher blocking rate \citep{TIBALDI1990}. Since we are using an extremely simple model of the atmosphere, it is also not surprising, that the blocking rates and lifetimes do not match quantitatively with observations. The lack of many dynamical and physical ingredients in our model is likely to be responsible for the fact that life times of blocking events are, to a good approximation, exponentially distributed, as opposed to the heavy tail properties found in \citet{Pelly2003}. They also define so called sector blocking which detects blocking which have a considerable size in the zonal direction. Given the rather coarse resolution of our model the observed blockings have at least an extent of roughly 1500 km which means that they already extent over a large area. Moreover, a discrimination of the events according to the blocking lengths (e.g. 2 - 3 days, 3 - 4 days, 4 - 5 days and 5 - 6 days) does not show any significant differences in the observed blocking rates. 

We conclude that despite the mentioned limitations the blocking index allows a meaningful definition of a blocked and an unblocked regime and that our model responds in at least qualitatively correct way to changes in the orography and meridional temperature gradient.  

\subsection{Linear Stability Of Blocking States}\label{sec:linearstability}
After having clarified in \cref{sec:blockingpatterns} that we indeed observe blocking events induced by orography, we will now evaluate the characteristics of the CLVs during blocked and unblocked phases. We follow up from the previous section and use the distinction between occurrence of blocking events and regular conditions to partition the attractor of the system, and then compute separately the statistical properties of CLVs and finite-time LEs in the two regions.

\begin{table*}[!ht]
\caption{Properties of the attractor without Orography}
\label{tab:chaoswooro}
\centering
\begin{tabular}{ccccc}
\toprule
$\Delta T$ [K]& Positive & Kaplan-Yorke & Metric & $1/\lambda_1$  \\
& Exponents & Dimension& Entropy [1/day]& [day]\\
\midrule
$39.81$ &$17$& $35.83$& $0.25$&$28.8$\\
$49.77$ &$55$& $125.82$& $3.15$&$6.8$\\
$66.36$ &$88$& $206.80$& $12.51$&$2.6$\\
$76.31$ &$98$& $232.09$& $18.66$&$1.9$\\
\bottomrule
\end{tabular}
\end{table*}
\begin{table*}[!ht]
\caption{Properties of the attractor with Orography}
\label{tab:chaosT1}
\centering
\begin{tabular}{cccccc}
\toprule
$\Delta T$ [K] & Height [km]& Positive & Kaplan-Yorke & Metric & $1/\lambda_1$\\
& & Exponents & Dimension& Entropy [1/day]&  [day]\\
\midrule
&$1.48$ &$18$& $38.11$& $0.29$&$26.8$\\
39.81 &$2.96$ &$19$& $41.51$& $0.35$&$24.5$\\
&$4.44$ &$19$& $41.24$& $0.34$&$24.5$\\
\toprule
&$1.48$ &$55$& $126.9$& $3.23$&$6.6$\\
49.77&$2.96$ &$55$& $129.6$& $3.42$&$6.4$\\
&$4.44$ &$56$& $131.6$& $3.56$&$6.2$\\
\toprule
&$1.48$ &$88$& $207.1$& $12.61$&$2.55$\\
66.36&$2.96$ &$88$& $207.5$& $12.72$&$2.54$\\
&$4.44$ &$89$& $207.9$& $12.85$&$2.51$\\
\toprule
&$1.48$ &$98$& $232.2$& $18.76$&$1.89$\\
76.31&$2.96$ &$99$& $232.2$& $18.82$&$1.89$\\
&$4.44$ &$99$& $232.4$& $18.88$&$1.88$\\
\bottomrule
\end{tabular}
\end{table*}

\begin{table*}[!ht]
\caption{Metric Entropy during the blocked and the unblocked phase. This is the sum of the in the long term averaged positive LEs of the Backward Lyapunov vectors averaged during the blocked and unblocked phases, respectively. \label{tab:chaosT3}}
\centering
\begin{tabular}{ccccc}
\toprule
$\Delta T$ [K] & Height [km]& Metric Entropy & Metric Entropy& Difference\\
& &  during blocking [1/day]& no blocking [1/day]&[1/day]\\
\midrule
      &$1.48$ & - & 0.29& - \\
39.81 &$2.96$ & - & 0.34& - \\
      &$4.44$ & - & 0.34& - \\
\toprule
      &$1.48$ & 3.41 & 3.21&0.20\\
49.77 &$2.96$ & 3.60 & 3.41&0.19\\
      &$4.44$ & 3.76 & 3.53&0.23\\
\toprule
      &$1.48$ & 12.71 & 12.54&0.17\\
66.36 &$2.96$ & 12.82 & 12.70&0.12\\
      &$4.44$ & 12.95 & 12.77&0.18\\
\toprule
      &$1.48$ & 18.87 & 18.67&0.20\\
76.31 &$2.96$ & 18.89 & 18.74&0.15\\
      &$4.44$ & 18.97 & 18.78&0.19\\
\bottomrule
\end{tabular}
\end{table*}
Let us start by examining basic dynamical and geometrical properties of the system. 
Starting from the  LEs we can derive the 
Kaplan-Yorke dimension and the metric entropy, see \cref{tab:chaosT1,tab:chaoswooro}. We assume, as often implicitly done, that the chaotic hypothesis \citet{Gallavotti1995a} holds. Therefore, we assume that our system has Axiom A-like properties and, in particular, that one has a Sinai-Ruelle-Bowen measure supported on its attractor and describing its asymptotic statistical and dynamical properties. The Kaplan-Yorke dimension is defined as
$
D_{KY}= k + \frac{\sum_{i=1}^k \lambda_i}{|\lambda_{k+1}|},
$
where $k$ is chosen so that the sum of the first $k$ LEs is positive and the sum of the first $k+1$ is negative.  This dimension is an upper bound of the fractal dimension of the attractor of the system. The metric entropy is given by the sum of the positive LEs and  a measure for the information creation \citep{Eckmann1985}. With increasing $\Delta T$ the Kaplan Yorke dimension and the metric entropy grow monotonically \citep{Lucarini2007}. While the observed motions are indeed chaotic for the studied values of $\Delta T$ and $h_0$, we can clearly see from these dynamical indicators that turbulence is much better developed for higher values of $\Delta T$. The impact of the orography on these numbers is small but it shows a small upward trend for larger $h_0$. This property shows that predicting weather becomes more complicated, if orography is added, since the characteristic predictability time decreases. We take the inverse of the leading LE $\lambda_1$ (in \cref{tab:chaosT1,tab:chaoswooro}) as a rough measure for predictability, since the rapid divergence of nearby trajectories is a necessary ingredient for having a fast error growth.
Note that different dynamical indicators are better suited to study the actual predictability of a system.  A better evaluation of the time scales of the system and the associated predictability could be obtained by studying systematically finite time/finite size LEs and the related  multifractal properties \citep{Boffetta1998,Boffetta2003}, which is outside the scope of this paper.
\begin{figure*}[h!]
\centering
 \includegraphics[width=\textwidth]{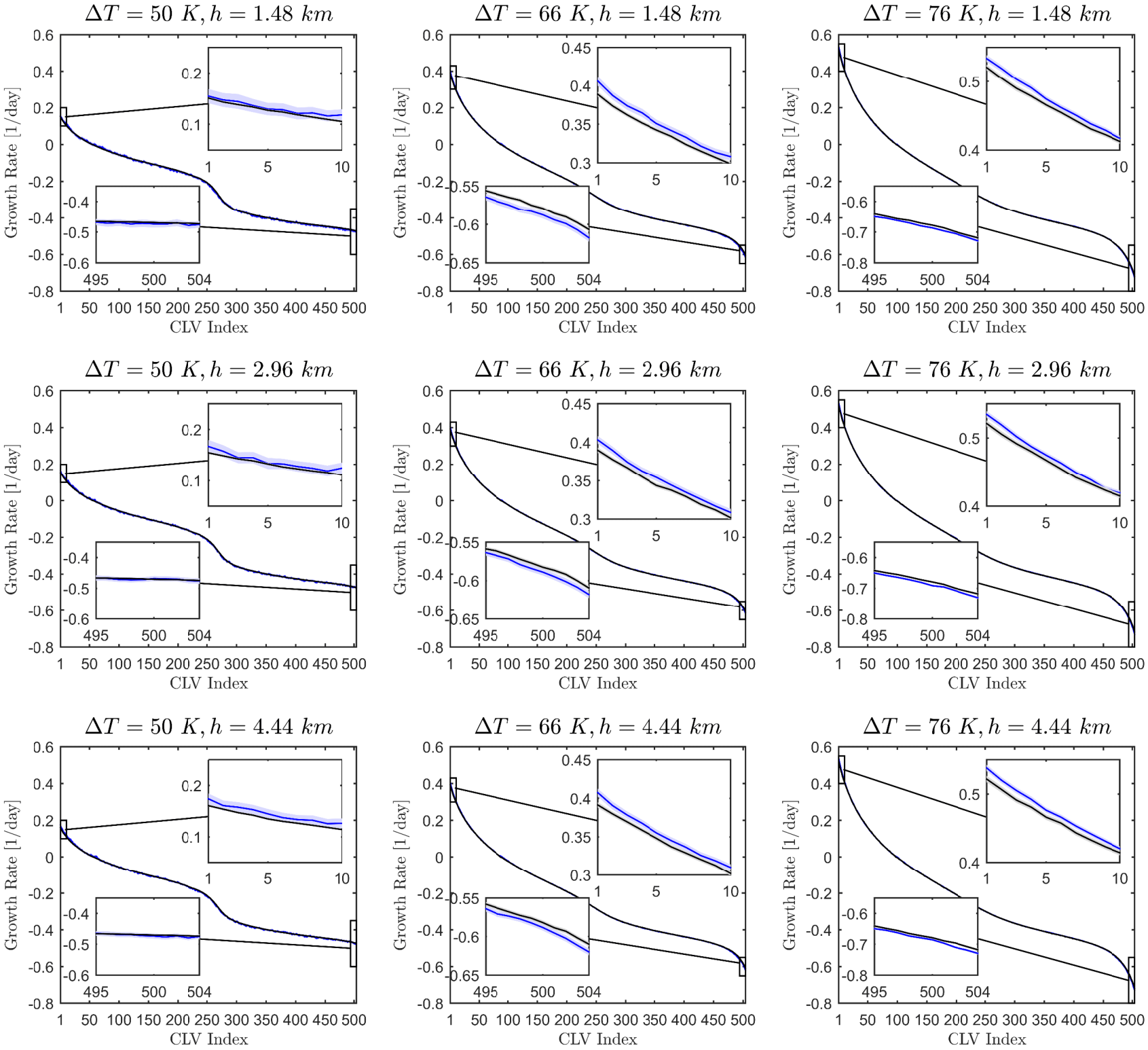}
 \caption{For the nine setups where we observe blocking (see \cref{fig:blockingrate}), the figure shows the differences in the growth rates during blocking (in blue color) versus unblocked phases (in black color). We additionally show the 3 $\sigma$ bars of confidence estimated by computing the degrees of freedom of the time series (shaded areas). For $\Delta T = 66 K, \, 76 K$ we can clearly estimate that the CLVs with highest/lowest LEs the baroclinic conversion increases/decreases significantly. For $\Delta T = 50 K$ such a tendency can not be clearly verified.\label{fig:lyapspectra}}
\end{figure*}

Let us now turn the focus on the average growth rates during the blocked phases and the unblocked phases. Since blocking conditions are relatively rare, at all practical levels this corresponds to comparing the finite time LEs computed during the blocked phase to the actual  long-term LEs. The growth rates of the ten leading CLVs increase significantly during blocking for the two largest $\Delta T$. The ten fastest decaying CLVs have significantly higher decay rates during the blocked phase.  This result also holds if only blockings with a certain length are considered (e.g. 2 - 3 days, 3 - 4 days, 4 - 5 days and 5 - 6 days). The statistical significance is determined by considering the 3 $\sigma$ confidence interval which is obtained by computing the degrees of freedom for each time series of the unblocked and blocked growth rates. The degrees of freedom is the number of effective observations of the averages over a fixed width which are not correlated with each other. This width is determined by computing the e-folding time of the autocorrelation of the time series, see \citet{Leith1973} and \citet{Mudelsee2010}. This indicates that the unstable CLVs grow globally faster. The question remains whether this is due to changes in the CLVs near or far away from the blocked region. This question will be partly answered in \cref{sec:localization} where we will be looking at spatial patterns.

The presence of larger positive LEs during the blocked phases indicate lower predictability. This seems in contradiction with the common knowledge that it is easier to predict the weather during blocking events. One can reconcile these two facts by considering that i) what we consider here is a global measure of predictability, not strictly relate to forecast skill near the region of the blocking; and ii) while predictability is higher during blocking events, it is extremely difficult to predict the onset and the decay of the blocking. Possibly, our result is related to the difficulty of capturing the regime transition. Let us refine a bit our mathematical statements. It is important to recall that predictability is usually characterized by the evolution of phase space volumes, hence the angles between the CLVs have to be considered as well. It is not necessary to actually use the CLVs for answering these questions, since the long term evolution of phase space volumes in the tangent linear regime is given by the the Backward Lyapunov Vectors (also called Gram-Schmidt vectors, see \citet{Kuptsov2012,Schubert2015}) which grow on average with the LEs, but their finite size LEs are different from the CLVs. In order to characterize the growth of a volume which covers all unstable directions, we use the fluctuations of the BLV-LEs (LEs of the Backward Lyapunov Vectors) to compute the metric entropy which is the sum of all positive LEs \citep{Eckmann1985}. In order to discriminate between the blocked and unblocked phase, we then average separately over these two phases of the flow (see \cref{tab:chaosT3}). The metric entropy confirms the slightly increased instability observed in the finite time LEs of the CLVs during blocking. We wish to remark that while considering larger values of $h_0$ leads to more frequent blocking events, no significant effect is instead found on the average growth rate of the disturbances. Orography plays an important role as a catalyzer for blocking events, more than influencing substantially their properties once they are realized. \subsection{Lorenz energy cycle during blocking}
\label{sec:lec}

\begin{figure*}[h!]
\includegraphics[width=\textwidth]{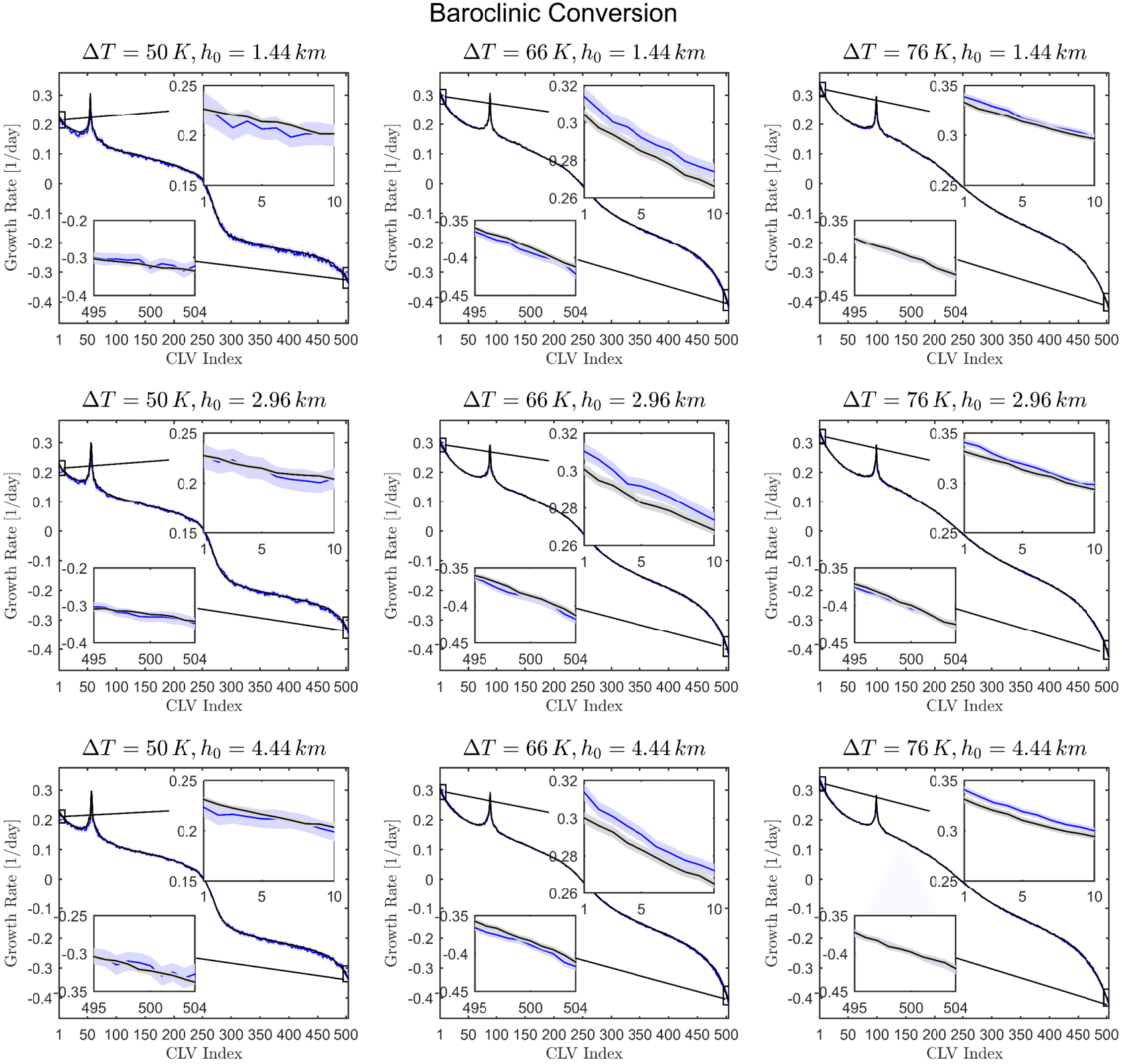}
\caption{For the nine setups where we observe blocking (see \cref{fig:blockingrate}), the figure compares the baroclinic conversion $\mathcal{C}_{BC}$ during blocking (in blue color) versus unblocked phases (in black color). We additionally show the 3 $\sigma$ bars of confidence estimated by computing the degrees of freedom of the time series (shaded areas). For $\Delta T = 66 K, \, 76 K$ we can clearly estimate that for the fastest growing CLVs, the baroclinic conversion increases significantly. For the fastest decaying a negative tendency can be observed, but with weaker statistical significance. For $\Delta T = 50 K$ such a tendency can not be clearly verified.\label{fig:bc}}
\end{figure*}
\begin{figure*}[h!]
\includegraphics[width=\textwidth]{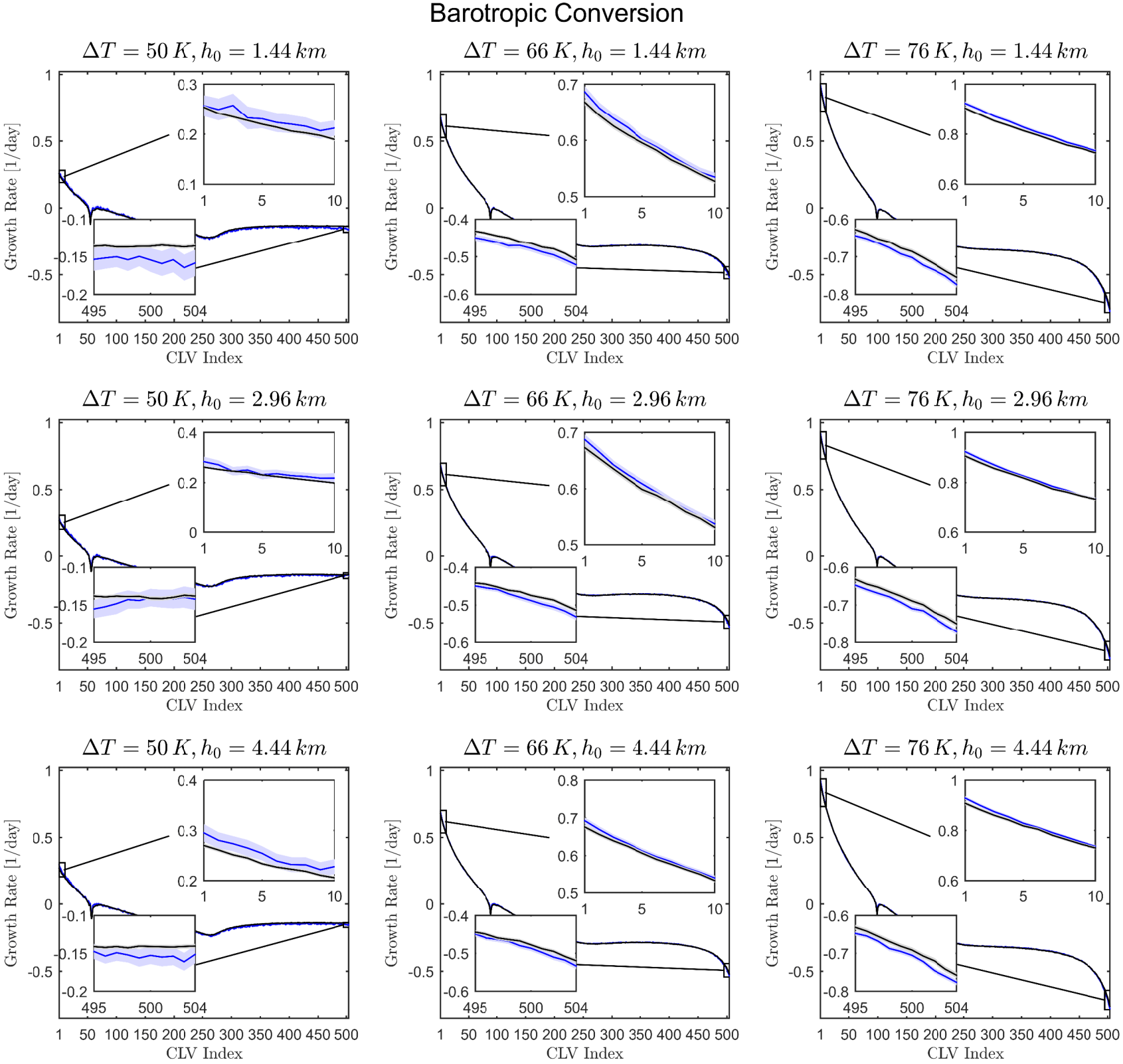}
\caption{For the nine setups where we observe blocking (see \cref{fig:blockingrate}), the figure shows the differences in the barotropic conversion $\mathcal{C}_{BT}$ during blocking (in blue color) versus unblocked phases (in black color). We additionally show the 3 $\sigma$ bars of confidence estimated by computing the degrees of freedom of the time series (shaded areas). For $\Delta T = 66 K, \, 76 K$ we can clearly estimate that for the CLVs with the highest/lowest LEs, the barotropic conversion increases/decreases significantly. For $\Delta T = 50 K$ such a tendency can not be clearly verified.\label{fig:bt}}
\end{figure*}
\begin{figure*}[h!]
\includegraphics[width=\textwidth]{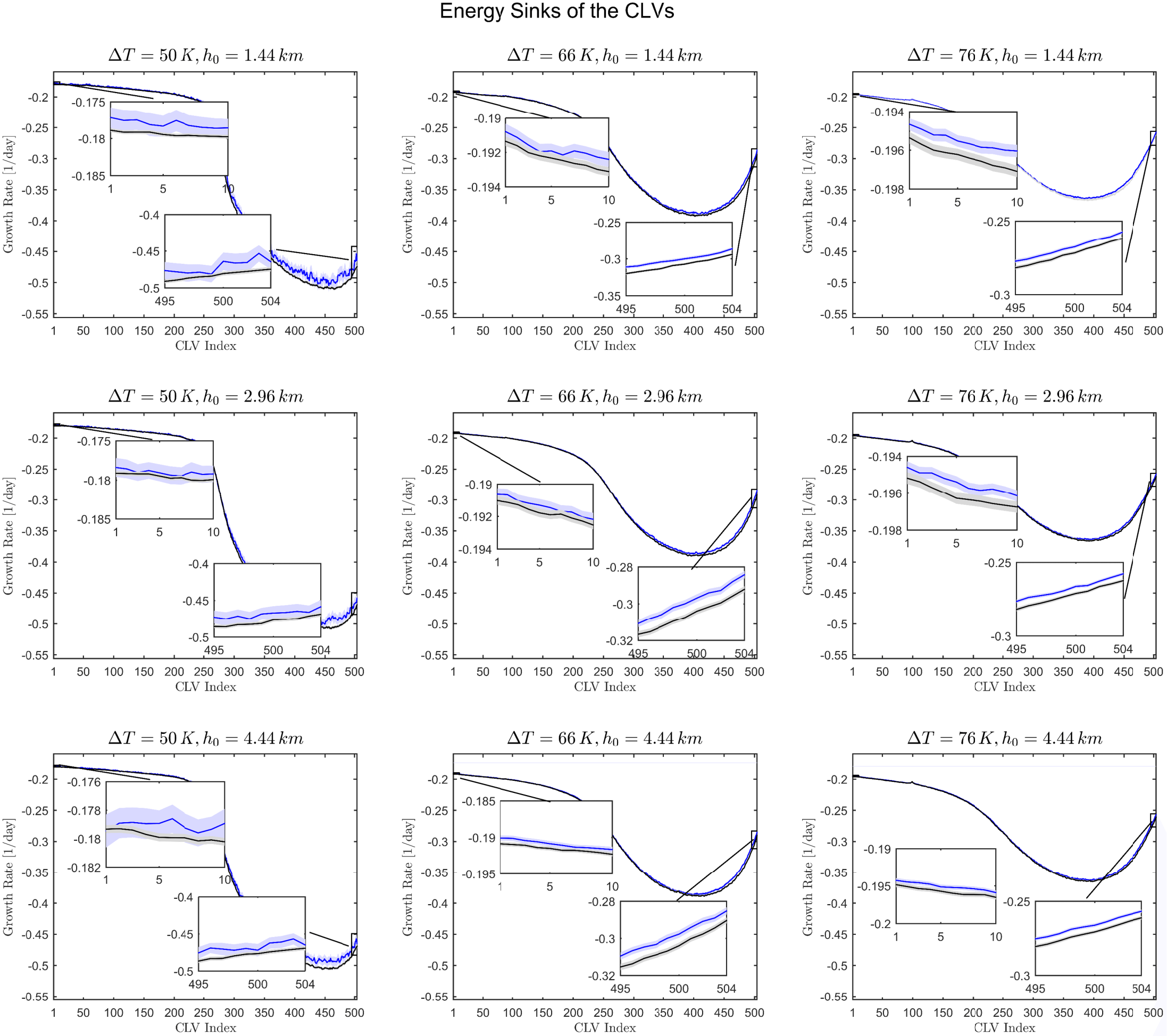}
\caption{The energy losses of the CLVs are the sum of the Ekman Friction, the heat and eddy diffusivity and newtonian cooling (see \citep{Schubert2015}). For the nine setups where we observe blocking (see \cref{fig:blockingrate}), the figure shows the sum of these terms during blocking (in blue color) versus unblocked phases (in black color). We additionally show the 3 $\sigma$ bars of confidence estimated by computing the degrees of freedom of the time series (shaded areas). For all $\Delta T $ we can clearly estimate that for the CLVs with the highest/lowest LEs, the energy losses decrease significantly.\label{fig:output}}
\end{figure*}

We want to give a physical interpretation to the lower predictability found during blocking events by studying the energetics of CLVs. In a previous work \citep{Schubert2015}, we made use of the fact that the CLVs are covariant solutions of the tangent linear equation and explained the growth and decay rate of the CLVs by looking at their Lorenz energy cycle. We showed that it is possible to study each individual term responsible for   energy conversions and sinks derived from the tangent linear equations. For details on that approach we would like to refer to this work. As opposed to the usual analysis of the energetics of linear perturbations of stationary background states,   in the case of the CLVs the background is fluctuating and the interactions between the perturbations and the background are more complex.  Therefore,
let us now briefly summarize how this energy cycle is computed. For each CLV, its average growth rate of the ($L^2$) square norm coincides with the average growth rate of its energy, thanks to the equivalence of norms in finite dimensional spaces. Hence, we can give a physical interpretation of the changes in the growth rates in the phases where blocking is present versus regular conditions (see \cref{sec:blockingpatterns}) by examining the details of the Lorenz energy cycle. We focus here on the budget of the total energy of the $j^{th}$ CLV $(\psi'_{1,j},\psi'_{2,j})$ resulting from the interaction with the background state  $\left(\psi^B_1,\psi^B_2\right)$ and from dissipative processes. For ease of notation, we will use for both fields also the barotropic streamfunction $\psi_P=\frac{\psi_1+\psi_2}{2}$ and the baroclinic streamfunction $\psi_T=\frac{\psi_1-\psi_2}{2}$  as well as  the respective geostrophic velocities $\textbf{v}=\left(u,v\right)=\left(-\partial_y\psi,\partial_x \psi\right)$:  
\begin{align}\label{eq:etot}
\begin{split}
 \ddt  E_{tot} =&\int d\sigma\bigg[\Delta\psi'_{1,j} \textbf{v}_1'\cdot\nabla \psi^B_1 -k_h\left( \psi'_{1,j} \Delta^2\psi'_{1,j}\right)\\&+\left< 1 \leftrightarrow 2 \right>+ 2 r \psi'_{2,j}\Delta\psi'_{2,j}\\
&\left.-\frac{2}{S}\psi'_{T,j}\textbf{v}_{P,j}'\cdot\nabla \psi^B_T+2 \frac{\kappa}{S} \psi'_{T,j}\Delta\psi'_{T,j}-2 \frac{r_R}{S} \psi_{T,j}^{'2}\right]
\end{split}
\end{align}
Since the CLVs are growing/decaying perturbations, we will normalize all the following energy conversion terms and sinks by the total energy of the considered CLV. We can group the various terms in \cref{eq:etot} into the barotropic conversion $\mathcal{C}_{BT}$ which increases to the kinetic energy of the CLVs and the baroclinic conversion $\mathcal{C}_{BC}$ which increases to the potential energy of the CLVs. Hence, both conversions contribute to the total energy of the CLVs. For our investigation, we will sum up all external forcings (newtonian cooling, diffusion and friction) acting on either the potential or kinetic energy in the term S. The energy budget for the total energy is then the following.
\begin{align}\label{eq:etot_budget}
 \ddt  E_{tot} = \mathcal{C}_{BC} + \mathcal{C}_{BT} +S
\end{align}
The baroclinic and barotropic energy conversion terms are defined as 
\begin{equation}\label{eq:cbc}
 \mathcal{C}_{BC}=\int d\sigma\left[-\frac{2}{S}\psi'_{T,j}\textbf{v}_{P,j}'\cdot\nabla \psi^B_T\right]
\end{equation}
and \begin{equation}\label{eq:cbt}
\begin{split}
 \mathcal{C}_{BT}=\int d\sigma\left[\Delta\psi'_{1,j} \textbf{v}_{1,j}'\cdot\nabla \psi^B_1+\Delta\psi'_{2,j} \textbf{v}'_{2,j}\cdot\nabla \psi^B_2\right],
\end{split}
\end{equation}
respectively. The energy loss is the sum of newtonian cooling, Ekman friction, and eddy diffusivity.
\begin{align}
\begin{split}
S=&\int d\sigma\left[-2k_h\left( \psi'_{T,j} \Delta^2\psi'_{T,j}+\psi'_{P,j} \Delta^2\psi'_{P,j}\right)\right.\\
&\left.+\ r \psi'_{2,j}\Delta\psi'_{2,j}+2 \frac{\kappa}{S} \psi'_{T,j}\Delta\psi'_{T,j}-2 \frac{r_R}{S} \psi_{T,j}^{'2}\right]
\end{split}
\end{align}
A positive average value for the baroclinic term $\mathcal{C_{BC}}$ implies that available potential energy of the background flow is converted into available potential energy of the $j^{th}$ CLV. The corresponding thermal fluctuations are then converted into kinetic energy of the CLV. Instead, an average positive value of the barotropic term $\mathcal{C_{BT}}$ implies a direct transfer of kinetic energy from the background flow to the $j^{th}$ CLV. Just like in the standard case of linear perturbations to a stationary background flow, also in this general scenario we have that a positive rate of baroclinic (barotropic) energy conversion rate is associated to a heat (momentum) flux opposite to the temperature (zonal momentum) gradient of the background flow. Such a negative feedback ensures the global stability of the system and is a manifestation of the second law of thermodynamics \citep{Schubert2015}. Clearly, it is a necessary condition for the LE corresponding to a CLV to be positive that at least one of the two terms $\mathcal{C_{BC}}$ or $\mathcal{C_{BT}}$ to be positive on the average. 

We can now study how the baroclinic and barotropic energy conversion rates are influenced by the presence of blocked flow conditions. The results are shown in \cref{fig:bc,fig:bt} for barotropic and baroclinic processes, respectively.   For completeness, we have also plotted the result for the energy sinks of the CLVs in \cref{fig:output}. 

For all analyzed configurations and for all CLVs, blocked conditions support smaller rates of energy dissipation than regular conditions. Nonetheless, such changes are numerically rather small and can be disregarded in the following discussion. 

In the case of weak baroclinic forcing ($\Delta T = 50 K$), the difference in the energetics of the CLVs between blocked and normal conditions is borderline or not statistically significant for most CLVs. Despite lack of strong statistical evidence, some useful indications can be given. Looking at the unstable CLV, we observe that during blocked phases the baroclinic conversion is lower than in usual conditions, whereas the opposite holds for the barotropic conversion. Therefore, we have that the (modest) enhanced growth rate of the unstable CLVs observed in blocked conditions (see \cref{fig:lyapspectra}) can be attributed to a more efficient barotropic conversion. Looking at the most stable CLVs, the situation is reversed, with barotropic (baroclinic) conversion rates being reduced (increased) in blocked conditions.  

The situation changes when considering conditions where stronger baroclinic forcing is imposed on the system ($\Delta T = 66 K$ and $\Delta T = 76 K$). We have that blocked conditions are accompanied by stronger baroclinic and stronger barotropic conversion rates for the unstable CLVs, while, conversely, the both conversion rates are reduced when looking at the most stable CLVs. 

These results seems to suggest that the energetics of blocking events is fundamentally different in background states featuring weak versus strong Equator to Pole temperature differences. In the former case, blocking is eminently related to modifications to the barotropic  instability of the flow, while in the latter case, it results from modifications of both barotropic and baroclinic instabilities. The synergy between the two forms of instability is likely to be responsible for the increase in the number of blocking events for larger values of $\Delta T$. Note that a strong sensitivity of the properties of the low-frequency variability on the intensity of the jet was already envisioned in \citet{Benzi1986} and verified by \citet{Ruti2006}.

The properties of the blocked states in terms of the Lorenz energy cycle of the CLVs are weakly dependent on the value of the perturbation orography $h_0$, which confirms in physical terms the eminently catalyzing role of orography for blocking.

\subsection{Localization Of CLVs}
\label{sec:localization}
We study now the localization of the CLVs during blocking conditions and compare it with what observed in normal unblocked phases.  A measure for the localization is given by the temporal variance of the CLVs at the grid points on the domain. In the control runs without orography, the variance of the CLVs does not depend on x. Results by  \citet{Szendro2008} suggest that if the zonal symmetry is broken due to orography, the CLVs will be localized in the x direction. 
\begin{figure}[h!]
     \centering
     \subfloat[][The first CLV]{\includegraphics[width=\textwidth]{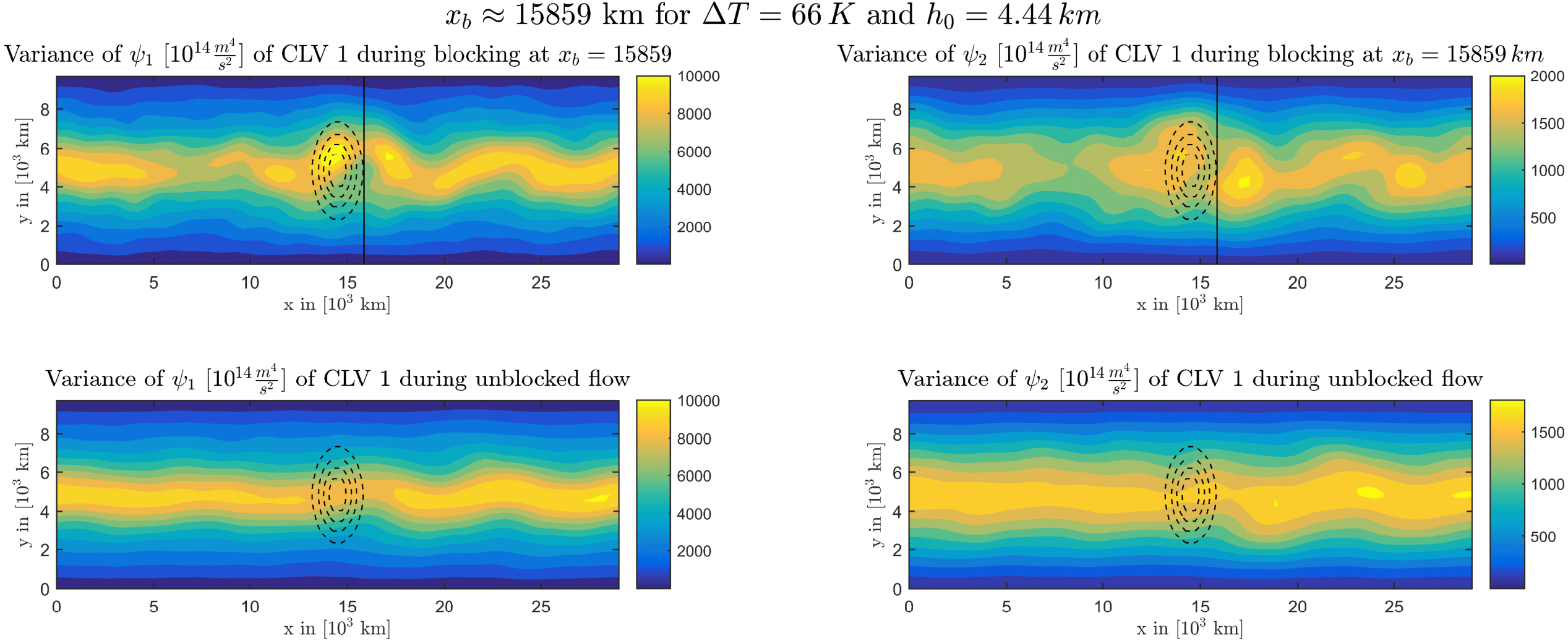}\label{fig:clv1}}\\
     \subfloat[][The $10^{th}$ CLV]{\includegraphics[width=\textwidth]{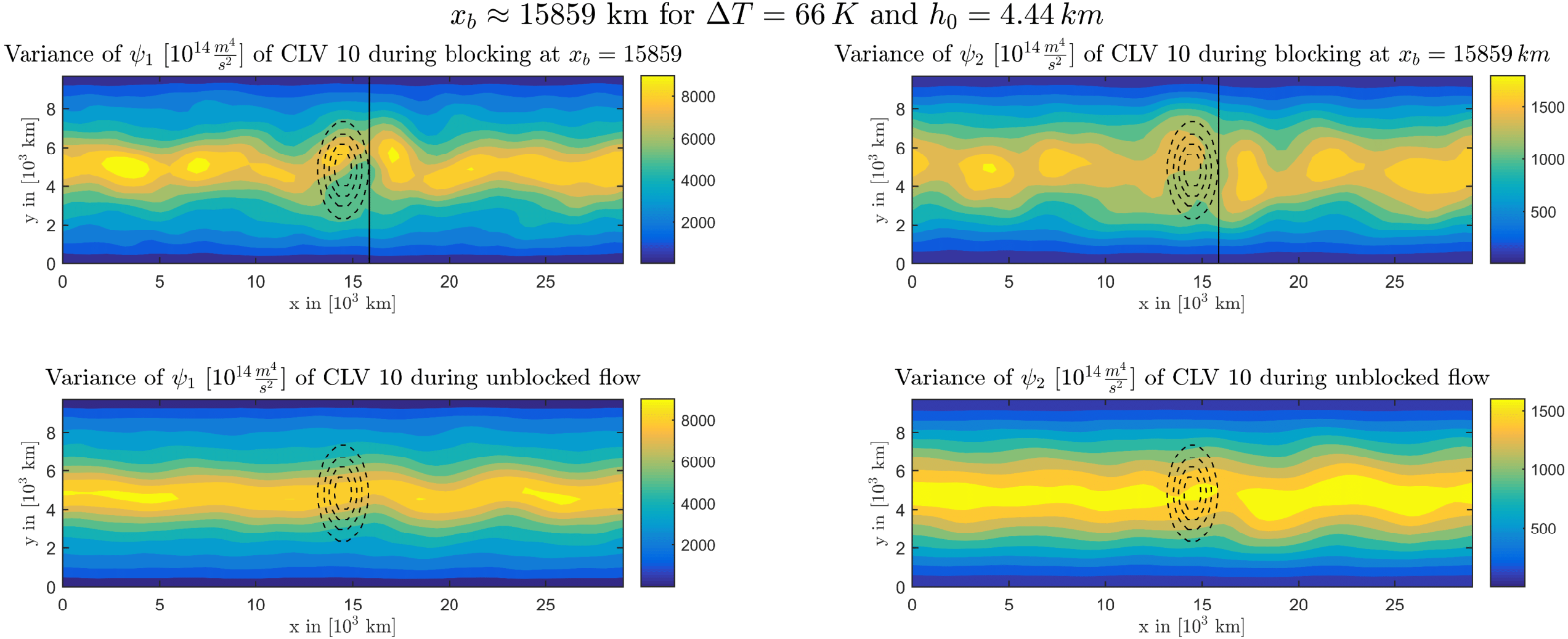}\label{fig:clv10}}\\
     \subfloat[][The $100^{th}$ CLV]{\includegraphics[width=\textwidth]{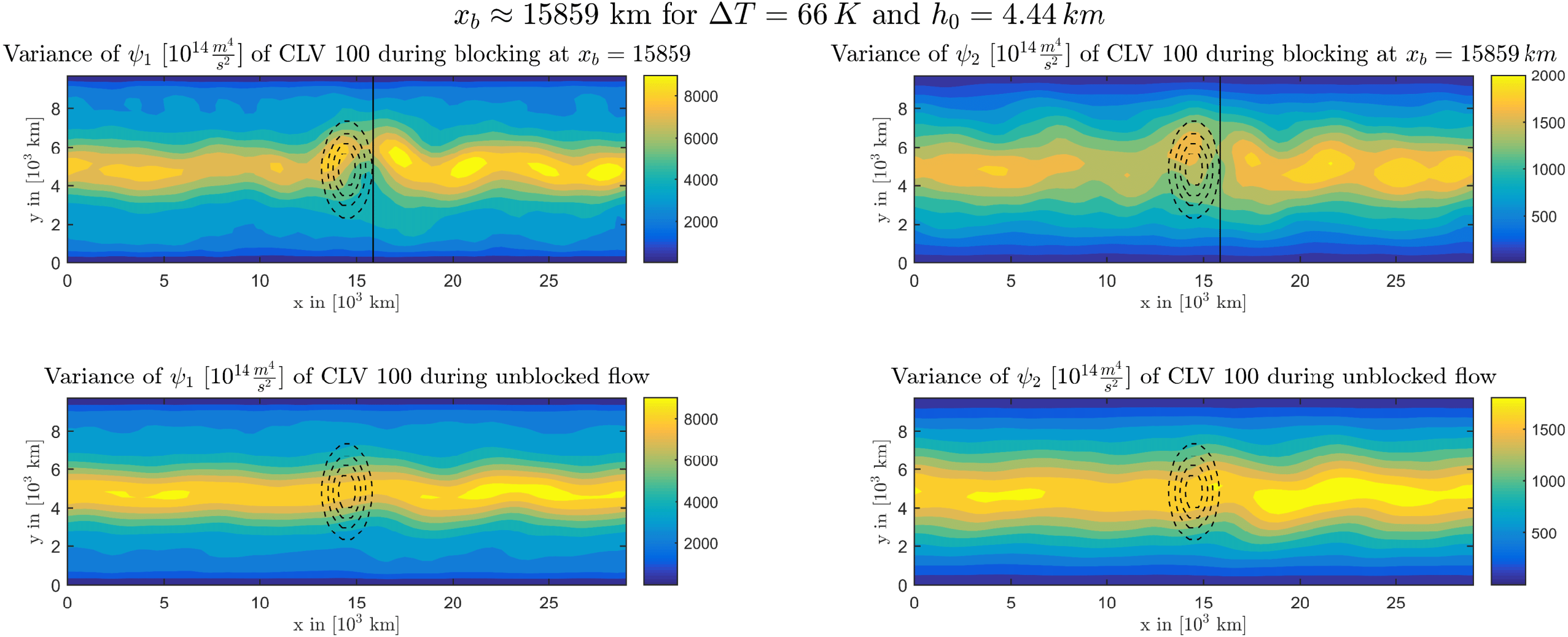}\label{fig:clv100}}
     \caption{Variance of selected covariant Lyapunov vectors. Panel (a) shows the first CLV, (b) the $10^{th}$ and Panel (c) shows the $100^{th}$ Panel. The left side of the Panels shows the upper layer stream function $\psi_1$ an d the right side shows the upper layer stream function $\psi_2$. The upper panels show the average for blocking at $x_b=15859\, km$. The lower panels show the average over the unblocked phase. The vertical black line shows the location of the blocked coordinate, whereas the dashed lines show the location of orography.  \label{fig:varianceclv}}
\end{figure}
In \cref{fig:varianceclv}, we show as an example the variance of three CLVs in the blocked phase and the unblocked phase for the blockings shown previously in \cref{fig:block35} (for  $\Delta T=66\, K$ and $h_0=4.44\, km$ and blocking at $x_{b}=15859\ km$). The figure shows a clear impact of the blocking  on the variance of the CLVs. Overall, the variance is localized in the meridional direction due to the symmetry break associated with the boundary conditions at   $y=0,\pi$, but at the location of the blocking the variance is shifted northward. In the zonal direction, the variance has a weak x-dependence (even if we have no reasons to expect zonal symmetry) during the unblocked phase. Away from the blocking, we see again a non zonal  disturbance with wave number four in both phases  (see \cref{sec:blockingpatterns}). Note that at the location of the block in the background state (see \cref{fig:block35}), the CLVs show a minimum of the variance. 

 Before we discuss the implications of these results, we would like to analyze the variance for all\ CLVs by slightly reducing the complexity of the data. We average the variance along the meridional direction and focus on the x-dependence only.
 Furthermore, we do not wish to analyze the average localization of the CLVs, but instead track the variations of the localization. Hence, we compute the ratio of the meridional average of the variance of the streamfunction of the CLVs (upper and lower layer, indicated by the upper indices 1 and 2, respectively) during blocking at a particular $x_b$ 
$$\sigma^{(1/2)}_{x_b}(x)=\frac{1}{\left|T_{x_b}\right|}\int_{t \in T_{x_b}} dt\int dy \left(\psi_{1/2}(x,y,t)^2- \left<\psi_{1/2}\right>^2 \right)$$and during the unblocked phase 
$$\sigma^{(1/2)}_{unbl}(x)=\frac{1}{\left|T_{unbl}\right|}\int_{t \in T_{unbl}} dt\int dy \left(\psi_{1/2}(x,y,t)^2- \left<\psi_{1/2}\right>^2 \right).$$
Note that $\left<\cdots\right>$ is the average along the whole available time series of the CLVs. We define the set $T_{x_b}$ to contain all time steps where the flow is blocked at $x_b$ and $T_{unbl}$ to contain all time steps which are not blocked. $\left|T\right|$ is the length of the respective phases. For this calculation, we transform the spectral fields into a [64x32] grid point field. Hence, we can view the $\sigma$s as the variance of the grid point amplitudes of the streamfunction during a blocking situation and during non-blocked phases. 

\begin{figure*}[h!]
\centering
 \includegraphics[width=\textwidth]{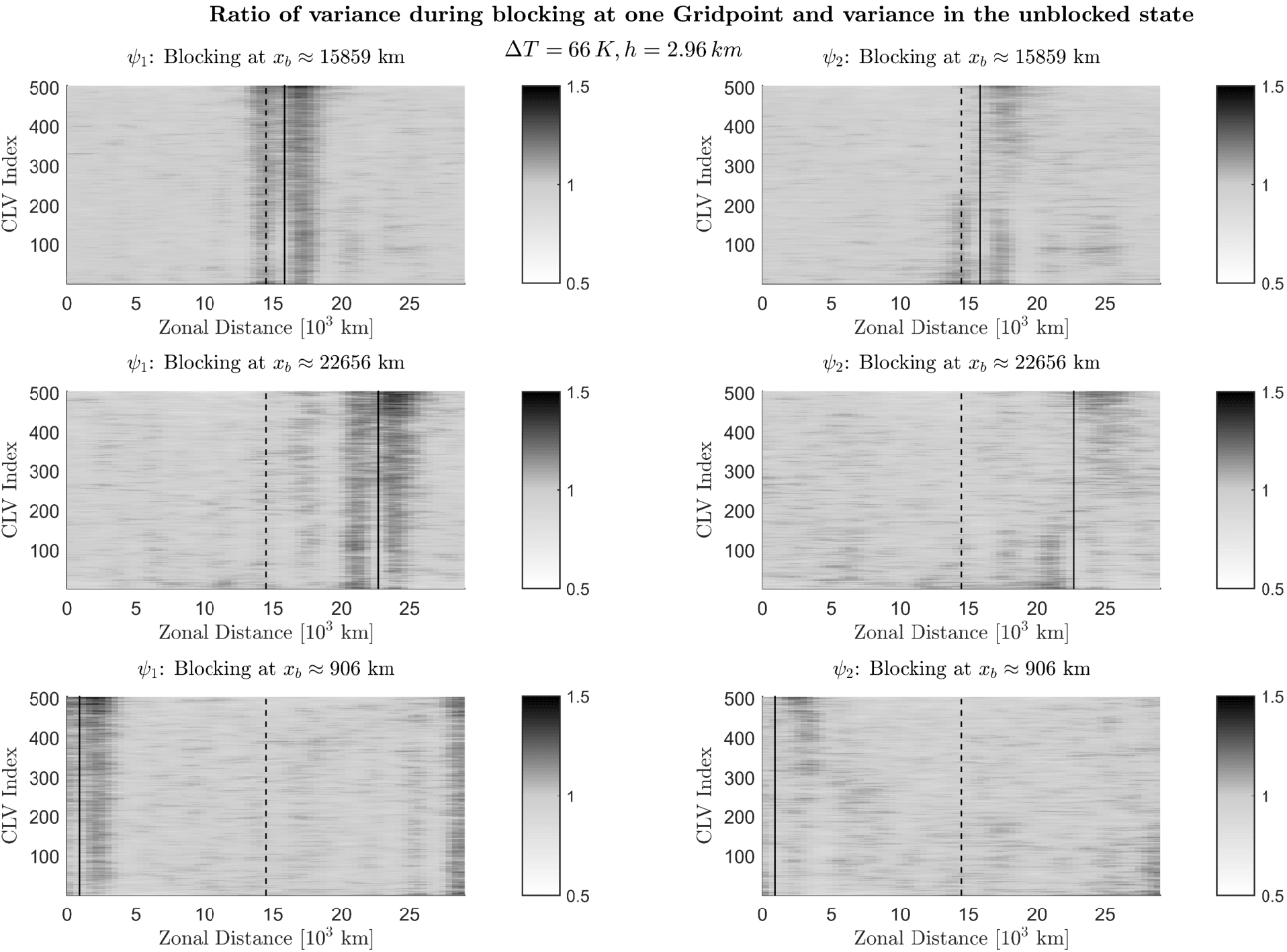}
 \caption{Since the localization of the CLVs is similar for the different setups with blocking events, we show here only results for $\Delta T = 66 K$ and $h=4.44\ km$. For every CLV (y-axis) we show the quotient of the meridionally averaged variance during blocking at a particular x coordinate (vertical solid black lines) and unblocked phases ($\Delta \mathcal{L}$ in \cref{eq:variance}).}\label{fig:variance}
\end{figure*}

We the measure the change in the localization  $\Delta \mathcal{L}$ as follows: 
\begin{equation}\label{eq:variance}\displaystyle
\Delta \mathcal{L}^{1/2}(x)=\frac{\sigma^{1/2}_{x_b}(x)}{\sigma^{1/2}_{unbl}(x)}
\end{equation}
If $\Delta \mathcal{L} > 1$, then a higher activity of the CLV (y-axis) during blocking at a particular zonal coordinate (x-axis) is implied. As an example, \cref{fig:variance} shows the results of the above \cref{eq:variance} for  the three local maxima of the blocking rate for $\Delta T=66\, K$ and $h_0=4.44\, km$ (see \cref{fig:blockingrate}). The other setups show similar results. In the figure, the vertical dashed line indicates the position of the peak of orography along the channel. We  present results for the the upper layer streamfunction $\psi_1$ (left panel) and the lower layer streamfunction $\psi_2$ (right panel). We see that the activity of almost all CLVs is higher close to the blocking and lower in the rest of the channel. To be more precise, the CLVs cluster around a region where blocking is detected. Note that the clustering occurs almost regardless of growth rate of the CLVs.
Also, the localization in the lower layer is less strong, which explains the reduction in the dissipation during blocked phases discussed before.
The variance of the CLVs at the center of the blocking is lower. This indicates that stability is higher in the center of the blocking compared to its borders. In order to clarify unambiguously this point,  the adjoint CLVs would have to be considered. These allow for projecting in a meaningful way an arbitrary  perturbation onto the non-orthogonal basis given by the CLVs. For a given time horizon, one could then obtain a characteristic growth/decay rate of linear perturbations in an arbitrary region of the flow. 

Note that using orthogonal Lyapunov vectors for this analysis would change the results. We can illustrate this by considering the example of optimal perturbations by \citet{Buizza1996}. Here, only the first optimal perturbation localizes close to the blocking, since this vector converges for long optimization times towards the first CLV, the remaining optimal perturbations can not behave in this way (see \cref{sec:CLV}). The results obtained using the physically relevant CLV basis underline that a transition to a blocked state is a change in the flow regime which effects all time scales and  processes. Hence, the linear dynamics of blocking events can not be reduced to a small number of changes on certain time scales and consequently, the detection of blocking events should take this into account.

\section{Summary and Conclusions}\label{sec:conclusion}
The goal of this paper is to study blocking events in a very simplified atmospheric model of the mid-latitudes. Blocking events are persistent deviations of the jet stream in the mid-latitudes from the usual quasi zonal symmetry.   Naturally, blocked states  possess very unusual properties in terms of weather forecast, and it is especially difficult to predicting the onset and decay of blocking events. It is well known that orography plays a major role in fostering the occurrence of blocking. 
This model is a quasi geostrophic two layer
model on a periodic channel with a beta plane approximation and driven by a forced meridional temperature gradient $\Delta T$. In the spirit of previous analysis of blocking, we added an orographic
forcing in order to produce enhanced blocking events in the flow \citep{Charney1980}. As orography, we use a Gaussian bump placed in the middle of the channel. We investigate four different values of $\Delta T$  (40 K, 50 K, 66 K and 76 K) in order to assess different degrees of large scale turbulence. The impact of orography is investigated with three different heights $h_0$  (1.48 km, 2.96 km and 4.44 km).

While such a setting is definitely outdated and insufficient in terms of providing a realistic statistics of blocking events, it provides qualitatively meaningful results and contains some of the essential physical and mathematical ingredients we want to consider, namely the possibility of having a turbulent state featuring a convincing Lorenz energy cycle fuelled by barotropic and baroclinic instabilities and damped by a variety of dissipative effects.

The main plus of such a simple model is that we are able to construct the CLVs, which are the covariant unstable and stable modes of the turbulent flow and provide a physical representation of the natural fluctuations of the flow. CLVs provide a complete descriptions of the dynamics and geometry of the attractor of the system and are useful for providing a new characterization of the properties of blocked vs regular conditions. 
 CLVs are the suitable choice for such an investigation, since they are covariant and independent of a norm. More precisely, they describe a first order approximation nearby trajectories and they evolve following local features of the non-linear background flow. Our objective in this study is to make a first step towards identifying the signature of the blocking events in the CLVs.

The model we consider here is was originally developed in order to analyze the long term averaged properties of CLVs and their capability to explain the variability of the full non linear flow \citep{Schubert2015}. There, we were solely interested in the long term average of the Lorenz energy cycle of the CLVs and to connect is to their growth or decay rates and property of being unstable or stable, thus linking the physics and the mathematics of the disturbances. The fluctuations of the CLVs or different "weather regimes" in the background state like blocking were not investigated. 

In the analysis presented here, we detect the blocking events with an Tibaldi-Molteni scheme \citep{TIBALDI1990}. Given the simplicity of the model we use, it is not surprising that the statistics of the events we label as blocking are only in qualitative agreement with what found in observations for all configurations we consider. Nevertheless, we can show that the detected events are indeed blocking highs which divert the jet stream from its zonal symmetry.  In the\ unblocked phase, it is also comforting to see that, the flow is more zonally symmetric and its mean state exhibits topographic Rossby waves. For higher meridional temperature gradients $\Delta T, $ the occurrence of the blocking rates increases, accompanied by a modest increase of its life time. The orography creates localized regions of high blocking rates. Such regions are located downstream of the orographic disturbance and their prominence is more evident when higher mountains are considered. The orographic influence is weaker when adopting a stronger baroclinic forcing.

Each CLV is associated to a LE, which measures its average growth (for unstable CLVs) or decay (for stable CLVs) rate. We have analyzed separately the growth rate of the various CLVs during the blocked and regular regimes.  Furthermore, the spatial variance during the blocked and unblocked phases is used for the localization of the CLVs.
Our results show a significant increase of the growth rate of the leading CLVs during blocked phases. Hence, the flow is more unstable during blocking. This might be interpreted as a trade off effect between increased stability in the blocked regions and less instability elsewhere in the flow. The increased instability indicates also that the regime transition to and from the blocked regime is in general difficult to predict. 
It should be noted that persistence as we expect it in the case of blocking and stability are not necessarily dependent on each other. We find support for our conclusions in recent results by  \citet{Faranda2015} which show that blockings can be connected to an unstable fixpoint in a reduced phase space.
In a closely related piece of work, Faranda et al.\footnote{These results are so far unpublished, yet they have been presented at the summer school  Statistical and mathematical tools for the study of climate extremes, held in Cargese, France, November 2015 and at the 2015 AGU assembly; see D. Faranda, P. Yiou and M. Carmen Alvarez Castro "Butterflies, Black swans and Dragon kings: How to use the Dynamical Systems Theory to build a "zoology" of mid-latitude circulation atmospheric extremes?" \url{https://agu.confex.com/agu/fm15/meetingapp.cgi/Paper/72315}} have shown that the dimensionality of the reduced phase space determining the dynamics of blocking events is higher than corresponding to regular quasi-zonal dynamics by exploiting a connection between extreme value statistics and the local dimension of the dynamical systems underlying attractor.  The fact that we not only observe increased instability in the CLVs but also an increased metric entropy during blockings fits very well to these results.

We have complemented the analysis of the instabilities by investigating the Lorenz energy cycle of the various CLVs and looking at the baroclinic and barotropic conversion rates. The enhancement of the growth rate in the blocked phase for the leading unstable CLVs is due to a strengthening of both barotropic and baroclinic conversion rates for intermediate and high values of $\Delta T$. Instead, for low values of $\Delta T$, enhanced instability of the unstable CLVs during blocked phase results from an enhancement of barotropic instability only. This clarifies that the dynamical processes behind blocking events are not the same in conditions of low versus high baroclinicty of the background flow. 

The variance of every CLV clusters around the blocked area. This hints for an increased instability at the boundaries of the detected blocking events. Instead, the spatial variance of the unstable CLVs is lower at the center of the blocking, possibly indicating higher local predictability.
The analysis via CLVs seems to provide a powerful way to link the occurrence of blocking events to specific modifications in the temporal and spatial properties of unstable processes responsible for energy conversion.

\citet{Gallavotti2014a} suggests that in multiscale systems it should be possible to associate the different spatial spatial and temporal scales of motion (e.g. macroscopic, mesoscopic and microscopic) to specific subsets of CLVs and related LEs. In particular, one expects that highly localized (extended) unstable CLVs might be associated to large (small) growth rates resulting from local (global) instabilities. 

It is tempting to follow this idea for investigating the complex portfolio of meteorological instabilities, but one needs indeed to consider more complex models than that adopted here. In particular, in a primitive equation model one could see whether it is possible to recognize small scale CLVs with high growth rate associated to mesoscale instabilities, while, additionally, CLVs associated to convective events should be found when non-hydrostatic models are adopted. 

Clearly, a QG model like the one used in this study is not appropriate for investigating these interesting aspects. Instead, the fact that here, as already observed in \citet{Schubert2015}, all CLVs have similar spatial scale (and the LEs, apart from those very close to zero, are also similar) is a "a posteriori" confirmation of the self-consistency of the scale analysis leading to the QG approximation. Another possibility is, instead, to use the formalism of CLVs and associated LEs to study a fluid model encompassing regions with different inertia and thermal inertia, like in the case of a coupled atmosphere-ocean model, in order to define rigorously, e.g., coupled modes of variability. 
\citet{Vannitsem2015} have recently provided extremely encouraging results in this direction using a severely truncated coupled atmosphere-ocean model. 
 
Future work should approach the characterization of blockings via CLVs in a model featuring spherical geometry and a realistic forcing, e.g. in \citet{Vannitsem2001},  in order to assess more persistent blockings.
Another direction of work is to define a local stability map of the flow using the adjoint CLVs. These would allow for quantifying the local growth rates of arbitrary linear perturbations to the flow. In this way, it should be possible to unambiguously identify local persistent structures like blocking.

\section*{Acknowledgement}
The authors wish to thank O. Talagrand, J. M. Lopez, G. Gallavotti, C. Franzke and F. Lunkeit for helpful discussions on the manuscript. We wish to thank as well T. Frisius for allowing to use and expand his QG model. V.L. acknowledges fundings from the Cluster of Excellence for Integrated Climate Science (CLISAP) and from the European Research Council under the European Community's Seventh Framework Programme (FP7/2007-2013)/ERC Grant agreement No. 257106. S. Schubert acknowledges fundings from the "International Max Planck Research School - Earth System Modeling" (IMPRS - ESM).

\bibliographystyle{plainnat}
\bibliography{library}
\end{document}